\documentclass[runningheads]{llncs}

\usepackage{arxiv_style}

\usepackage{graphicx}
\usepackage{booktabs}

\usepackage[accsupp]{axessibility}  %

\usepackage[hidelinks]{hyperref}

\usepackage{orcidlink}
\usepackage{pifont}

\usepackage{comment}
\usepackage{enumitem}
\usepackage{multirow}
\usepackage{wrapfig}
\usepackage{colortbl}
\usepackage[normalem]{ulem}

\usepackage{caption}
\usepackage{graphicx}
\usepackage{booktabs}

\setlist[itemize]{align=parleft,left=0pt}

\definecolor{kwon}{rgb}{0.36, 0.54, 0.66}
\definecolor{kjs}{rgb}{0.784, 0.003, 0.313}
\definecolor{azure(colorwheel)}{rgb}{0.0, 0.5, 1.0}
\definecolor{nicegreen}{rgb}{0.0, 0.7, 0.1}
\definecolor{ashblue}{rgb}{0.36, 0.54, 0.66}
\definecolor{ashgrey}{rgb}{0.7, 0.75, 0.71}
\definecolor{applegreen}{rgb}{0.55, 0.71, 0.0}
\definecolor{jy}{rgb}{0.58, 0, 0.827}
\definecolor{cornellred}{rgb}{0.7, 0.11, 0.11}
\definecolor{darkcyan}{rgb}{0.0, 0.55, 0.55}
\definecolor{CuGray}{gray}{0.9}
\definecolor{airforceblue}{rgb}{0.36, 0.54, 0.66}
\definecolor{rev}{rgb}{0.784, 0.003, 0.313}
\definecolor{pink}{cmyk}{0, 0.7808, 0.4429, 0.1412}
\definecolor{amethyst}{rgb}{0.6, 0.4, 0.8}
\definecolor{black}{rgb}{0.0, 0.0, 0.0}
\definecolor{tb3_yellow}{rgb}{0.996, 1.0, 0.6}
\definecolor{tb3_orange}{rgb}{0.980, 0.8, 0.604}
\definecolor{tb3_red}{rgb}{0.972, 0.6, 0.6}
\definecolor{dimgray}{rgb}{0.41, 0.41, 0.41}
\definecolor{brickred}{rgb}{0.8, 0.25, 0.33}
\definecolor{bleudefrance}{rgb}{0.19, 0.55, 0.91}
\definecolor{blue(ncs)}{rgb}{0.265, 0.445, 0.765}
\definecolor{cyan}{rgb}{0.0, 1.0, 1.0}
\definecolor{darkbrown}{rgb}{0.4, 0.26, 0.13}
\definecolor{brown(traditional)}{rgb}{0.59, 0.29, 0.0}
\definecolor{magenta}{rgb}{0.98, 0.176, 0.815}

\newcolumntype{g}{>{\columncolor{CuGray}}c}
\newcolumntype{z}{>{\columncolor{CuGray}}l}

\renewcommand{\paragraph}[1]{\vspace{1mm}\noindent\textbf{#1.}\,\,}

\usepackage{xspace}

\makeatletter
\def\@fnsymbol#1{\ensuremath{\ifcase#1\or *\or \dagger\or \ddagger\or
   \mathsection\or \mathparagraph\or \|\or **\or \dagger\dagger
   \or \ddagger\ddagger \else\@ctrerr\fi}}
\makeatother

\def\onedot{.\@\xspace}
\def\eg{\emph{e.g}\onedot} 
\def\ie{\emph{i.e}\onedot}

\def\etal{\emph{et al}\onedot}

\newcommand{\Sref}[1]{Sec.~\ref{#1}}
\newcommand{\Eref}[1]{Eq.~(\ref{#1})}

\def\rmI{{\mathbf{I}}}

\def\rmL{{\mathbf{L}}}

\newcommand{\Real}{\mathbb R}

\newcommand{\be}{\begin{eqnarray}}
\newcommand{\ee}{\end{eqnarray}}
\newcommand{\bee}{\begin{eqnarray*}}
\newcommand{\eee}{\end{eqnarray*}}

\newcommand{\matrixb}{\left[ \begin{array}}
\newcommand{\matrixe}{\end{array} \right]}

\definecolor{green(ncs)}{rgb}{0.0, 0.62, 0.42}

\DeclareMathOperator{\proj}{proj}
\newcommand{\vct}{\mathbf}
\newcommand{\vctproj}[2][]{\proj_{\vct{#1}}{#2}}

\begin{document}

\title{Learning-based Axial Video Motion Magnification}

\author{Kwon Byung-Ki\inst{1}\orcidlink{0000-0003-4187-7944} \ \ \
Oh Hyun-Bin\inst{2}\orcidlink{0009-0002-1511-7482} \ \ \
Kim Jun-Seong\inst{2}\orcidlink{0000-0001-7570-6508} \ \ \  \\
Hyunwoo Ha\inst{2}\orcidlink{0009-0005-3723-3451} \ \ \
Tae-Hyun Oh\inst{1,2,3}\orcidlink{0000-0003-0468-1571}}

\authorrunning{Byung-Ki ~\etal}

\institute{Graduate School of AI, POSTECH  \and
Department of Electrical Engineering, POSTECH \and
Institute for Convergence Research and Education in Advanced Technology, Yonsei University \\
\email{\{byungki.kwon, hyunbinoh, junseong.kim, hyunwooha, taehyun\}@postech.ac.kr}}

\maketitle

\begin{abstract}
Video motion magnification amplifies invisible small motions to be perceptible, which
provides humans with a spatially dense and holistic understanding of small motions in the scene of interest.
This is based on the premise that magnifying small motions enhances the legibility of motions.
In the real world, however, vibrating objects often possess
convoluted
systems that have complex natural frequencies, modes, and directions.
Existing motion magnification often fails to improve legibility since the intricate motions still retain complex characteristics even after
being magnified, which likely distracts us from analyzing them.
In this work, we focus on improving legibility by proposing a new concept, \emph{axial} video motion magnification, which magnifies decomposed motions along the user-specified direction.
Axial video motion magnification can be applied to various applications where motions of specific axes are critical, by providing simplified and easily readable motion information.
To achieve this,
we propose a novel
Motion Separation Module that enables the disentangling and magnifying of motion representation along axes of interest.
Furthermore, we build a new synthetic training dataset for our task that is generalized to real data.
Our proposed method improves the legibility of resulting motions along certain axes
by
adding a new feature: user controllability.
In addition, axial video motion magnification is a more generalized concept; thus,
our method can be directly adapted to the \emph{generic} motion magnification and achieves favorable performance against competing methods. The code and dataset are available on our project page: \url{https://axial-momag.github.io/axial-momag/}.
\end{abstract}

\section{Introduction}
\label{sec:intro}

Small motions often convey important signals in practical applications,
\eg, building structure health monitoring~\cite{cha2017output, chen2014structural, chen2017video, chen2015modal, chen2015developments, qiu2018defect}, machinery fault detection~\cite{sarrafi2018vibration, vernekar2014gear,smieja2021motion}, sound recovery~\cite{davis2014visual}, and healthcare~\cite{balakrishnan2013detecting, brattoli2021unsupervised, moya2020non, fan2021robotically, janatka2020surgical}.
Video motion magnification~\cite{liu2005motion,wadhwa2014riesz,oh2018learning,wadhwa2013phase,wu2012eulerian} is the technique to amplify subtle motions in a video, revealing details of motion that are hard to perceive with the naked eyes.
This allows users to grasp dense and holistic behavior information of the scene of interest instantly, as long as the resulting motion is simple and easily interpretable.
However, in practice, vibrating objects in the real world often possess complex systems that have complicated natural frequencies, modes, and directions~\cite{oliveto1997complex}.
Even after being magnified, the intricate movement persists, which restricts the advantages of motion magnification because the key underlying premise of its effectiveness is based on the legibility of the magnified motion in aforementioned applications, \ie, effectively understanding the way objects move.

\begin{figure}[t]
\centering
\includegraphics[width=0.93\linewidth]{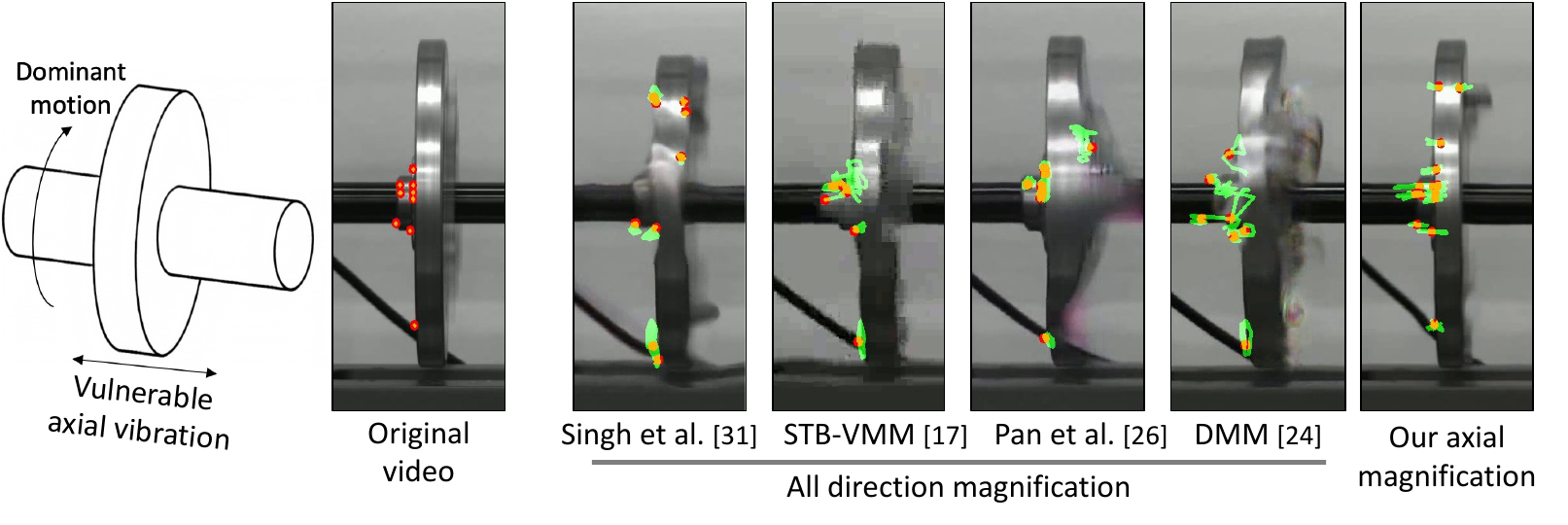}
    \caption{
    \textbf{Importance of axial motion magnification.}
    When identifying faults in rotating machinery, analysis of the vulnerable axial vibration is critical~\cite{yang2005vibex,luo2021analysis}. Existing learning-based methods~\cite{oh2018learning,lado2023stb,pan2024self,singh2023multi} amplify
    motions along all axes, which yield artifacts.
    It hinders the analyses of vulnerable axial vibration.
    This motivates the importance of our axial motion magnification that magnifies decomposed motions along a user-specified axis.
    We magnify the axial vibration only, achieving artifact-free results and the legibility of critical motions.
    For the visualization purpose, we overlay the sample trajectories obtained from the Kanade-Lucas-Tomasi (KLT) Tracker~\cite{lucas1981iterative}.
    }
    \label{figure:teaser}
\end{figure}

In this work, we focus on improving the legibility of magnified motion by proposing a novel concept, \emph{axial} video motion magnification, which magnifies decomposed motions along the user-specified direction.
All the existing works, \eg, ~\cite{wu2012eulerian,wadhwa2013phase,wadhwa2014riesz,oh2018learning,lado2023stb,singh2023multi,pan2024self}, have overlooked this key importance of legibility according to axes, although
there are many practical cases where the importance of motion varies by axes. In the fault detection application of machines, vibration direction serves as the key component for the expert tree of rotating machinery’s fault diagnosis~\cite{yang2005vibex}. As shown in Fig.~\ref{figure:teaser}, even small motions along the vulnerable axis are critical, while dominant rotational motions are not~\cite{luo2021analysis}.
Likewise, many apparatus consisting of natural or artificial materials often have
vulnerable axes due to the asymmetry property of microstructures, \eg, fracture toughness~\cite{tilbrook2006crack, brodnik2021fracture, li2023computational}.
This motivates us to separately analyze axial motions.

Specifically, we propose a novel learning-based axial video motion magnification method, where the motions in a user-specified axis are magnified.
Our method can independently magnify small motions along two orthogonal orientation axes with two independent magnification factors for each axis, which facilitates the analysis of complex small motions in the lens of axes favorable to the user.
To this end, we propose
the Motion Separation Module (MSM), which disentangles the motion representation into two orthogonal orientations and manipulates it in the direction specified by the user.
To train the proposed neural network, we develop and build a new synthetic dataset for the axial motion magnification task.
Therefore, our proposed approach adds a new user control feature, which improves
the legibility of resulting motions along a certain axis. This allows our axial motion magnification to become a generalization of the existing \emph{generic} motion magnification.
Thus, our method can be directly adopted to the generic motion magnification task and achieve favorable performance against competing methods.
We summarize our contributions as follows:
\begin{itemize}
    \setlength{\itemsep}{1pt}
    \item We propose the new concept, learning-based axial video motion magnification, which allows us to selectively amplify small motions along a specific direction.
    \item We propose and analyze the Motion Separation Module (MSM) for the axial motion magnification.
    We find that adopting MSM is effective not only in axial magnification but also in distinguishing small motions from noise.
    \item We propose a way to synthesize a new synthetic dataset to train the proposed axial motion magnification model and exhibit generalization to real data.
\end{itemize}

\section{Related Work}
\label{sec:RL}
Liu~\etal\cite{liu2005motion} first pioneered
the video motion magnification task, which involves
estimating explicit motion trajectory via optical flow to generate magnified frames.
They group and filter the motion trajectories based on motion similarity and user's intervention, and magnify them through explicit image warping, followed by video inpainting to fill holes created by the explicit warping.

Wu~\etal\cite{wu2012eulerian} re-formulate the motion magnification task as an Eulerian method that represents motion
by intensity changes of pixels at each fixed
location without actual movement~\cite{freeman1991motion}.
The Eulerian approach, \eg, \cite{wu2012eulerian,wadhwa2013phase,wadhwa2014riesz,zhang2017video,takeda2018jerk,oh2018learning, takeda2019video,takeda2020local,takeda2022bilateral}, becomes standard in motion magnification due to its noise robustness, sensitivity to small motions, and simple system by avoiding
challenging
warp and inpaint approach for filling holes and handling occlusions.
The system of the Eulerian methods typically consists of motion representation, manipulation, and reconstruction. The previous works can be categorized into two main focuses: 1) proposing motion representations or 2) motion manipulation methods.

In the first category,  Wu~\etal~\cite{wu2012eulerian} present the motion representation motivated by the first-order Taylor expansion, which is implemented by the Laplacian pyramid as spatial decomposition.
Wadhwa~\etal~\cite{wadhwa2013phase,wadhwa2014riesz} enhance the representation by modeling the motion as phase representations, which are implemented by
complex steerable filters~\cite{simoncelli1995steerable} in \cite{wadhwa2013phase} and Riesz transform in \cite{wadhwa2014riesz} as spatial decomposition, respectively.
These works rely on
the classic signal processing theory with such hand-designed spatial filter designs,
which do not model non-linear phenomenons and yield artifacts and noisy results.

To deal with, Oh~\etal~\cite{oh2018learning} first coined learning-based video motion magnification, called Deep Motion Magnification (DMM), by modeling motion representation with deep neural networks.
As no real data exists for training video motion magnification, they propose a method to build motion magnification synthetic data.
With the development, other learning-based variants~\cite{lado2023stb,singh2023multi} have been proposed, focusing on neural network architectures.
These approaches demonstrate promising results by
effectively handling diverse challenging scenarios such as occlusion and noisy inputs.
Also, the motion magnification factors of the data-driven approaches can be controlled by the way the synthetic dataset is generated, while those of the traditional methods~\cite{wu2012eulerian,wadhwa2013phase} are theoretically restricted.

In the second category, when Wu~\etal~\cite{wu2012eulerian} present Eulerian motion magnification, they also propose to use a temporal filter on the motion representation to select the motion frequency of interest.
This allows the noise to be suppressed by focusing on specific motions and increasing the legibility of magnified motion.
There were attempts to extend to increase the legibility by proposing temporal filters to magnify different types of motions and deal with artifacts from large motions:
acceleration~\cite{zhang2017video,takeda2018jerk}, intensity-aware temporal filter~\cite{takeda2022bilateral}, velocity or all-frequency filter~\cite{oh2018learning}.
Our work is compatible with all these methods.

In this work, we present a new notion of motion magnification by disentangling motion axes of the user's interest.
We design a neural architecture to induce disentanglement of motion in oriented axes. Also, we propose the synthetic data generation pipeline for the axial motion magnification task.

\section{Learning-based Axial Motion Magnification}\label{mm}
We first discuss preliminaries about generic
motion magnification, which refers to the methods that amplify the motion regardless of direction, including the prior arts
\cite{wu2012eulerian, oh2018learning, singh2023multi,lado2023stb,pan2024self} (\Sref{gmm}).
Then, we re-frame the motion magnification problem in the view of axial motion magnification (\Sref{omm}), and elaborate on our network architecture, and synthetic data generation method (\Sref{nn}).
\subsection{Preliminary -- Generic Motion Magnification}\label{gmm}
Following the convention~\cite{wu2012eulerian, wadhwa2013phase}, for simplicity,
we consider
the 1D image intensity
being
shifted by the displacement function $\delta(x,t)$, which is parameterized by position $x$ and time $t$. It can be generalized to local translational motion in 2D image~\cite{wu2012eulerian}.
Given an underlying intensity profile function $f(\cdot)$,
the 1D
image intensity
$I(x,t)$ can be represented as
\begin{equation}\label{eq1}
I(x,t) = f(x + \delta(x,t)).
\end{equation}
The goal of motion magnification is to synthesize the magnified image
$\hat{I}(x,t)$:
\begin{equation}\label{eq:def_motionmag}
\hat{I}(x,t) = f(x + (1+\alpha)\delta(x,t)),
\end{equation}
where
$\alpha$ denotes the magnification factor.
The key factor of motion magnification methods
lies in the extraction of
the displacement function $\delta(x,t)$ from \Eref{eq1}.
If $\delta(x,t)$ can be decomposed,
we can approximate $\hat{I}(x,t)$ by multiplying $\delta(x,t)$ with the magnification factor $\alpha$ and applying the reverse of the decomposition process.
However, it is an ill-posed problem to extract exact displacements from the observed intensity images~\cite{wu2012eulerian}.
Therefore, the prior arts approximately decompose
$\delta(x,t)$; for example,
Wu~\etal~\cite{wu2012eulerian} use the first-order Taylor expansion as:
\begin{equation}
    I(x,t) \approx f(x) + \delta(x,t) \tfrac{\partial f(x)}{\partial x}.
\end{equation}
Learning-based methods~\cite{oh2018learning,lado2023stb,singh2023multi} design neural networks that have intermediate representations related to $\delta(\cdot)$, called shape representation. The representations are multiplied by $\alpha$, followed by reconstruction for magnification.

\subsection{Axial Motion Magnification}\label{omm}
To introduce the axial motion magnification task,
we now consider the 2D spatial coordinate by
slightly abusing the notations, \eg, $\textbf{x}=(x,y)$ to refer to the coordinate in the 2D image intensity $I(\textbf{x}, t)$.

\paragraph{Problem Definition}
We can represent
$I(\textbf{x}, t) = f(\textbf{x} + \boldsymbol{\delta}(\textbf{x}, t))$
with
a 2D displacement vector $\boldsymbol{\delta}(\textbf{x}, t) \in \Real^{2}$.
Given an angle $\phi\,{\in}\,\Real$ of the user-specified direction of interest, the goal of the
axial
motion magnification task is to isolate and amplify
the motion component corresponding to the direction angle $\phi$ within the displacement vector.
We represent the axially magnified image $\hat{I}^{\phi}(\textbf{x},t)$ as
\begin{equation}\label{eq2}
\hat{I}^{\phi}(\textbf{x},t) = f(\textbf{x}+\alpha^{\phi}\boldsymbol{\delta}^{\phi}(\textbf{x},t)),
\end{equation}
where $\alpha^{\phi} \geq 0$ denotes the axial magnification factor and $\boldsymbol{\delta}^{\phi}(\textbf{x},t)$
the projection of $\boldsymbol{\delta}(\textbf{x},t)$ onto a 2D directional unit vector $\textbf{p}^{\phi}$ with the angle $\phi$, \ie, the motion component.
We can break down the motion component $\boldsymbol{\delta}^{\phi}(\textbf{x},t)$ into:
\begin{equation}
    \boldsymbol{\delta}^{\phi}(\textbf{x},t) = \vctproj[\textbf{p}^{\phi}]{\boldsymbol{\delta}(\textbf{x},t)}.
\end{equation}

\paragraph{Relationship with Generic Motion Magnification}
If we obtain $\boldsymbol{\delta}(\textbf{x},t)$, we can determine $\boldsymbol{\delta}^{\phi}(\textbf{x},t)$ and $\boldsymbol{\delta}^{\phi_\perp}(\textbf{x},t)$ through the projections onto $\textbf{p}^{\phi}$ and $\textbf{p}^{\phi_\perp}$. In this case, we can extend Eq.~\ref{eq2} to represent not only the displacement vector of an angle $\boldsymbol{\delta}^{\phi}(\textbf{x},t)$ but also of its orthogonal direction $\boldsymbol{\delta}^{\phi_\perp}(\textbf{x},t)$, as
\begin{equation}\label{eq3}
\hat{I}^{\phi}(\textbf{x},t) = f(\textbf{x}+\alpha^{\phi}\boldsymbol{\delta}^{\phi}(\textbf{x},t)+\alpha^{\phi_\perp}\boldsymbol{\delta}^{\phi_\perp}(\textbf{x},t)),
\end{equation}
where $\alpha_{\phi}$, $\alpha_{\phi_\perp} \geq 0$ denotes the axial magnification factors corresponding to the $\phi$ and $\phi_\perp$ directions, respectively.
This formulation encompasses the various motion magnification scenarios, \eg, axial and generic motion magnifications. Setting $\alpha^{\phi_\perp}$ to $0$ leads to the formulation resulting in axial motion magnification, while setting $\alpha^{\phi}$ equal to $\alpha^{\phi_\perp}$ results in generic motion magnification.

\subsection{Neural Networks and Training}\label{nn}
Departing from the previous learning-based methods that are confined to generic motion magnification~\cite{oh2018learning,singh2023multi,lado2023stb,pan2024self}, we introduce a novel neural network architecture and a dedicated training dataset designed to learn two angle-aware motion representations proportional to the motion displacement $\boldsymbol{\delta}^{\phi}$ and $\boldsymbol{\delta}^{\phi_\perp}$, respectively.
These allow our approach to unveil a distinctive feature: the magnification of motion in user-defined directions while retaining the functionality for generic motion magnification.

\paragraph{Network Architecture}
\begin{figure}[t]
\centering
\includegraphics[width=1\textwidth]{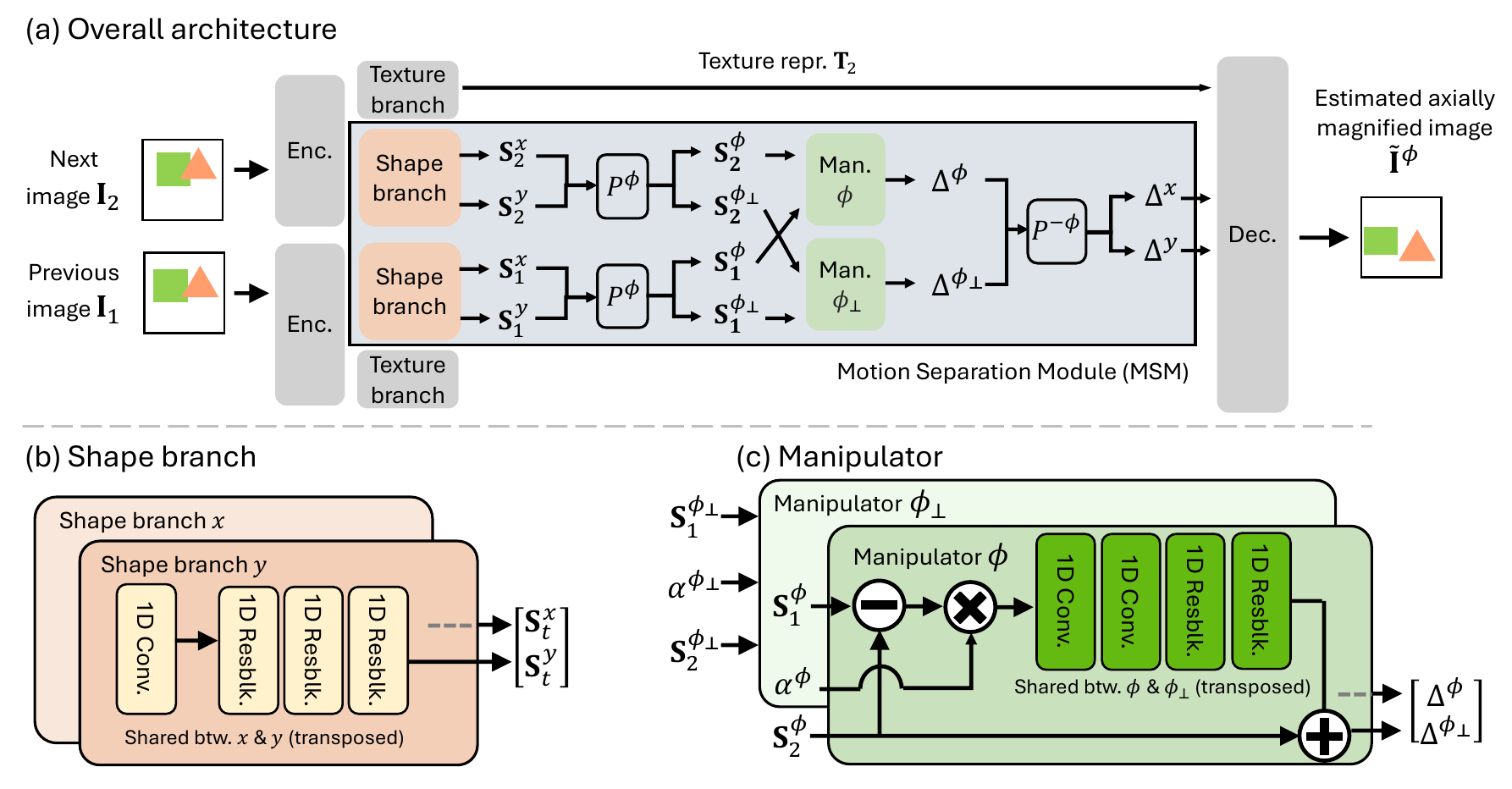}
    \caption{
    \textbf{Proposed architecture.}
    (a) The \emph{Encoder} outputs features from input images and the features are fed to the \emph{Texture} branch and Motion Separation Module (MSM). (b) Using weight-shared 1D convolutions, the \emph{Shape} branch extracts shape representations along the $x$ and $y$-axes. These representations are fed to the projection layer $P^{\phi}$, which generates axial shape representations, \ie, $\textbf{S}^{\phi}_{t}$ and $\textbf{S}^{\phi_\perp}_{t}$. (c) the \emph{Manipulator} amplifies
    them by the axial magnification factors and the inverse projection layer $P^{-\phi}$ re-project them onto the $x$ and $y$-axes.
    Finally, the \emph{Decoder} predicts the axially magnified image from the outputs from both the \emph{Texture} branch and MSM.
    }
    \label{figure:architecture}
\end{figure}

Our whole architecture consists of \emph{Encoder}, \emph{Texture} \& \emph{Shape} branches, \emph{Manipulator}, and \emph{Decoder} similar to DMM~\cite{oh2018learning} (see Fig.~\ref{figure:architecture}-(a)), where texture represents color and texture-related information while shape represents scene structure-related information that later leads to motion $\boldsymbol{\delta}$~\cite{oh2018learning}.
To extract axial shape representations,
we design Motion Separation Module (MSM) consisting of the completely re-designed and dedicated \emph{Shape} branch and \emph{Manipulator} as depicted in Fig.~\ref{figure:architecture}-(b,c).
In MSM, instead of extracting a single specified direction's $\boldsymbol{\delta}^{\phi}$, we design to extract its orthogonal direction's $\boldsymbol{\delta}^{\phi_\perp}$ as well. This design choice is motivated by the extended axial motion magnification equation Eq.~\ref{eq3} and enables conducting various motion magnifications, including both axial and generic motion magnifications.

Given consecutive input video frames $\textbf{I}_{t} \in \Real^{H\times W \times 3}$ at $t=1$ and $t=2$ for example,
texture representations $\textbf{T}_{t} \in \Real^{H/4\times W/4\times 32}$ are obtained by $\textbf{T}_{t}=F(E(\textbf{I}_{t}))$, where
$E(\cdot)$ and $F(\cdot)$ denote the \emph{Encoder} and the \emph{Texture} branch, respectively.
The outputs of $E$ are fed into MSM.
The same output from $E$ is fed into the \emph{Texture} branch and MSM, respectively.

To extract
the motion representations along two orthogonal orientations and manipulate them based on the user-defined angle, we grant the learnable parameters to learn the directionality in MSM.
Our \emph{Shape} branch $G(\cdot)$ first extracts the axial shape representations
along the canonical $x$ and $y$-axes by applying weight-shared 1D convolutions but with spatially transposing the convolution kernels, yielding
$[\textbf{S}^{x}_{t} ,\textbf{S}^{y}_{t}]$=$G(E(\textbf{I}_{t}))$ where $\textbf{S}^{x}_{t}, \textbf{S}^{y}_{t}
\in \Real^{H/2\times W/2\times 32}$.
Then, these
are projected by the \emph{projection} layer, which produces axial shape representations of $\phi$ and $\phi_\perp$ directions, \ie, $\textbf{S}^{\phi}_{t}$ and $\textbf{S}^{\phi_\perp}_{t}$.
Motivated by the steerable filters~\cite{freeman1991design}, where an arbitrarily rotated representation can be synthesized by a linear combination of directional representations, we design the projection layer $P^{\phi}$
with a linear matrix as
\begin{equation}
P^{\phi}\left(\left[\begin{array}{l}
S_t^x \\
S_t^y
\end{array}\right]\right)=\left[\begin{array}{cc}
\cos \phi & \sin \phi \\
-\sin \phi & \cos \phi
\end{array}\right]\left[\begin{array}{l}
S_t^x \\
S_t^y
\end{array}\right]=\left[\begin{array}{l}
S_t^\phi \\
S_t^{\phi_{\perp}}
\end{array}\right].
\end{equation}

\begin{figure}[t]
\centering
\includegraphics[width=1\textwidth]{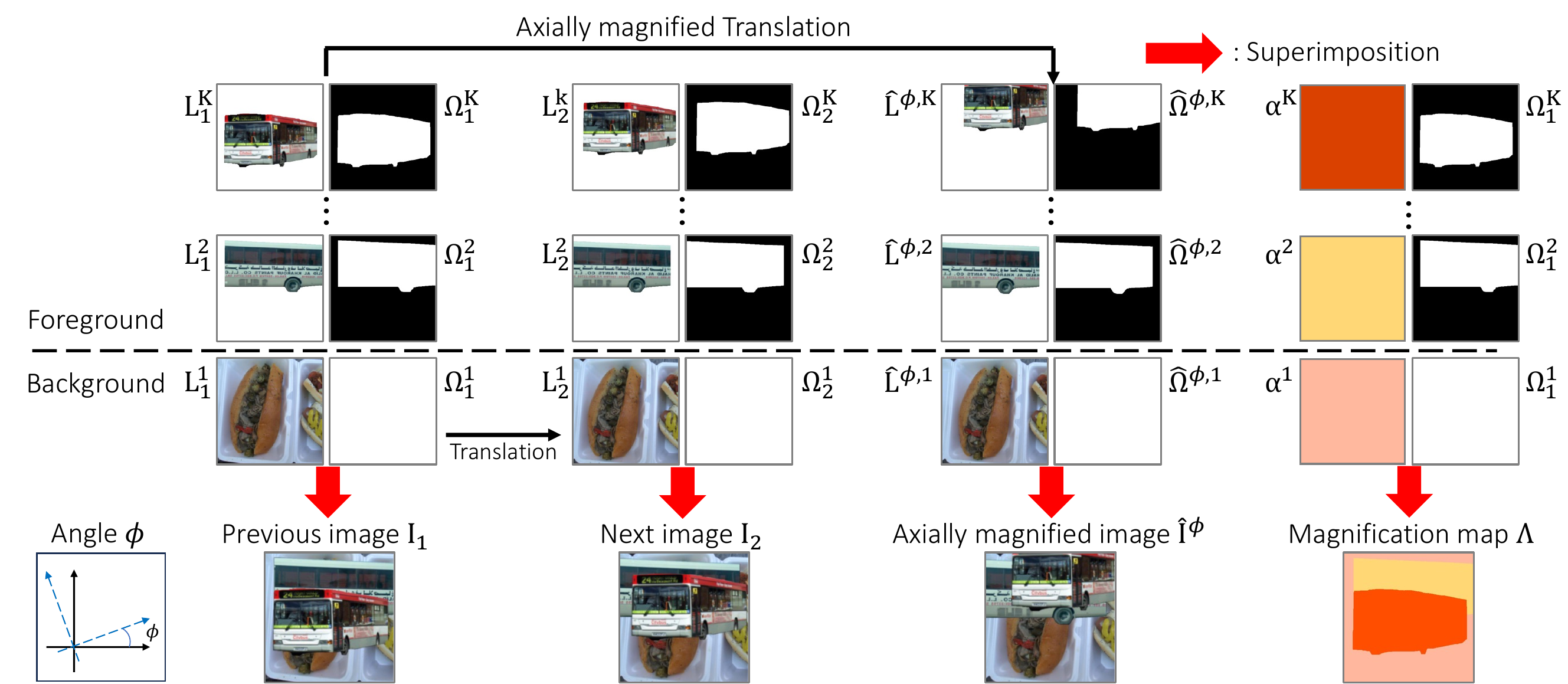}

    \caption{
    \textbf{Synthetic data generation pipeline for axial motion magnification.}
From the sampled background and foregrounds, each with their own segmentation masks, we compose the previous layer images $\{\rmL^k_{1}\}_{k=1}^{K}$ and masks $\{\boldsymbol{\Omega}^k_{1}\}_{k=1}^{K}$.
To generate next layer images $\{\rmL^k_{2}\}_{k=1}^{K}$ and masks $\{\boldsymbol{\Omega}^k_{2}\}_{k=1}^{K}$, we apply the random translations to $\{\rmL^k_{1}\}_{k=1}^{K}$ and $\{\boldsymbol{\Omega}^k_{1}\}_{k=1}^{K}$.
Axially magnified layer images $\{\hat{\rmL}^{\phi,k}\}_{k=1}^{K}$ and masks $\{\hat{\boldsymbol{\Omega}}^{\phi,k}\}_{k=1}^{K}$ are also synthesized by translations but with the axially magnified translation parameters.
These images and masks are then superimposed into a single image to yield $\textbf{I}_{1}$, $\textbf{I}_{2}$, and $\hat{\textbf{I}}^{\phi}$, respectively.
The dataset also include angles $\phi$ and the object-wise magnification maps $\boldsymbol{\Lambda}$ generated by superimposing $\{\boldsymbol{\alpha}^{k}\}_{k=1}^{K}$ with $\{\boldsymbol{\Omega}_{1}^k\}_{k=1}^{K}$.}
    \label{figure:datagen}
\end{figure}

The \emph{Manipulator} $M(\cdot)$ computes
the difference of the axial shape representations and magnifies them by multiplying the axial magnification factors $\alpha^{\phi}$.
Then, these manipulated representations are fed into subsequent 1D convolutions, and added to the axial shape representation $\textbf{S}^{\phi}_{2}$.
For $\phi_\perp$, we use the same manipulator, of which weights are shared but spatially transposed, for applying $\alpha^{\phi_\perp}$.
Note that, with this separation of $\phi$ and $\phi_\perp$,
we can set the magnification factors $\alpha^{\phi}$ and $\alpha^{\phi_\perp}$ independently, enabling broad applications of controls as another benefit.
For the outputs of the \emph{Manipulator} $\Delta^{\phi}, \Delta^{\phi_{\perp}}$, where $\Delta^{\phi}=M(\textbf{S}^{\phi}_{1}, \textbf{S}^{\phi}_{2}, \alpha^{\phi})$, we re-project them onto the canonical $x$ and $y$-axes by inverse projection layer $P^{-\phi}$, obtaining $\Delta^{x}, \Delta^{y}$.
Finally, the \emph{Decoder} $D(\cdot)$ predicts the axially magnified output frame $\tilde{\textbf{I}}^{\phi}$ as
\vspace{-2mm}
\begin{equation}
    \tilde{\textbf{I}}^{\phi} = D\left(\textbf{T}_{2}, \Delta^{x}, \Delta^{y}\right).
\end{equation}
This network architecture enables the network to conduct both generic and axial motion magnification, given the user setting of the angle $\phi$.
The model is trained with the loss function suggested by DMM~\cite{oh2018learning} with a slight modification to impose the loss separately to the x-axis and y-axis shape representations.
Details of the loss function can be found in the supplementary material.

\paragraph{Training Data Generation}
In the real world, acquiring consecutive images and magnified images at the same time is impossible.
Due to this,
DMM~\cite{oh2018learning} proposes a synthetic training dataset for the generic motion magnification task.
However, this dataset is not sufficient to induce the disentanglement of the axial property we need.
Thus, we propose a new synthetic dataset specifically designed for
the
axial
 motion magnification, where the motion between $\textbf{I}_{1}$ and $\hat{\textbf{I}}^{\phi}$ is associated with the angle $\phi$ and axial magnification factor vector
$\boldsymbol{\alpha}{=}(\alpha^{\phi}; \alpha^{\phi_\perp})$.
Motivated by the synthetic dataset generation protocol of DMM,
we synthesize the training data pairs using the widely adopted simple copy-paste method~\cite{oh2018learning,ghiasi2021simple}.

Figure~\ref{figure:datagen} shows the synthetic data generation pipeline.
We sample one background from COCO~\cite{COCO} and $K{-}1$ number of foreground textures with segmentation masks from PASCAL VOC~\cite{PASCAL}.
These elements are randomly located on image planes of resolution 384$\times$384 to produce $K$ previous layer images $\{\rmL^k_{1}\}_{k=1}^{K}$ and corresponding masks $\{\boldsymbol{\Omega}^k_{1}\}_{k=1}^{K}$.
Following this, with randomly sampled $K$ translation parameters
$\{ \textbf{d}^{k} \}_{k=1}^{K}$, we generate the next layer images $\{\rmL^k_{2}\}_{k=1}^{K}$ and masks $\{\boldsymbol{\Omega}^k_{2}\}_{k=1}^{K}$ by translating the initial layers and masks according to $\{ \textbf{d}^{k} \}_{k=1}^{K}$.
For the axially magnified layer images $\{\hat{\rmL}^{\phi,k}\}_{k=1}^{K}$ and their masks $\{\hat{\boldsymbol{\Omega}}^{\phi,k}\}_{k=1}^{K}$, we sample $K$ axial magnification vectors $\{\boldsymbol{\alpha}^{k}\}_{k=1}^{K}$ and a single degree of angle $\phi$.
Then, we perform the same procedure as the next layers but with the axially magnified translation parameters $\{ \boldsymbol{\alpha}^{k}(\vctproj[\textbf{p}^{\phi}]{\textbf{d}^{k}};\vctproj[\textbf{p}^{\phi_\perp}]{\textbf{d}^{k}}) \}_{k=1}^{K}$.
These previous, next, and axially magnified layer images and masks are then superimposed into a single image to yield $\textbf{I}_{1}$, $\textbf{I}_{2}$, and $\hat{\textbf{I}}^{\phi}$, respectively.
Our dataset also includes the angle $\phi$ and the object-wise magnification map $\boldsymbol{\Lambda}$ which is generated by superimposing $\{\boldsymbol{\alpha}^{k}\}_{k=1}^{K}$ segmented with $\{\boldsymbol{\Omega}_{1}^k\}_{k=1}^{K}$.
We observe that utilizing both $\phi$ and $\boldsymbol{\Lambda}$ are useful for learning the representations distinguishing small motions from noises, which will be discussed on~\Sref{ablation}. Additionally, the adaptation of both $\phi$ and $\boldsymbol{\Lambda}$ enables pixel-wise axial motion magnification.
We provide more details
in the supplementary materials.

\section{Experiments}
\paragraph{Implementation Details}
We train our learning-based axial motion magnification network on the newly proposed dataset, which contains a total of 100k samples, for 50 epochs with a batch size of $8$ and a learning rate $2\times 10^{-4}$. For training, we use two NVIDIA TITAN RTX GPUs.

\paragraph{Evaluation Setup}
We examine the performance of our method in axial and generic motion magnification, respectively.
In generic motion magnification, we compare our method to the phase-based method~\cite{wadhwa2013phase}, Singh~\etal~\cite{singh2023multi}, STB-VMM~\cite{lado2023stb}, Pan~\etal~\cite{pan2024self}, and DMM~\cite{oh2018learning}.
In axial motion magnification, there is no method of handling a user-specified angle and performing axial magnification due to our novel problem setup. Therefore, we propose a new axial baseline, called \textit{modified phase-based}, by modifying Wadhwa~\etal\cite{wadhwa2013phase}. Specifically, we modulate the phase-based method~\cite{wadhwa2013phase} to operate in the axial scenario by employing a half-octave bandwidth pyramid and two orientations, with one of them having its phase representation manipulated along the axis of interest.
Following DMM~\cite{oh2018learning}, we use both the \textit{dynamic} and $\textit{static}$ modes
in the experiments.
Additional experiments of diverse scenarios and implementation details can be found in the supplementary material and video.

\subsection{Axial Motion Magnification}
We evaluate our method compared to the modified phase-based
method
in the axial motion magnification scenario to demonstrate the effectiveness of the learning-based axial motion magnification.

\begin{figure*}[t]
\centering
\includegraphics[width=1\linewidth]{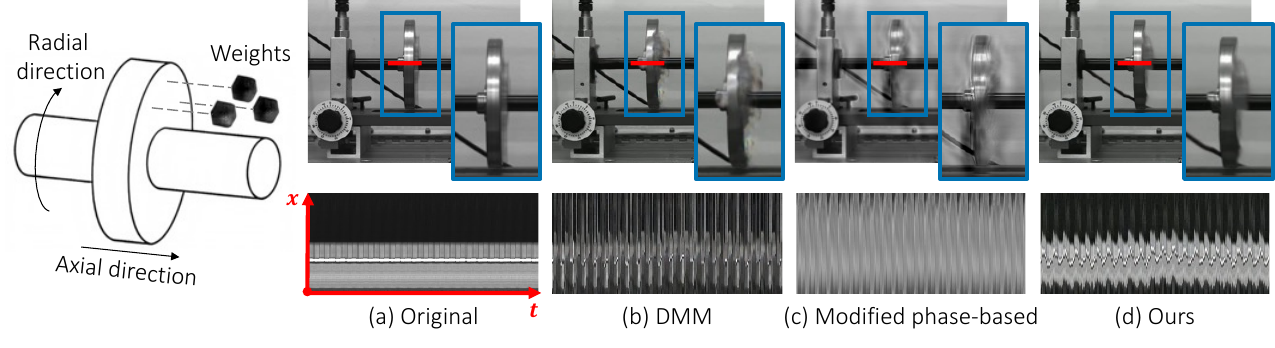}
    \caption{\textbf{[Left] Imposing an imbalance on a rotor, [Right] Qualitative results in axial motion magnification scenario.}
    We attach weights to a rotor to impose an imbalance and acquire \textit{rotor imbalance} sequence, which has axial vibrations.
    Then, we amplify only the motion of rotor's axial direction with the magnification factor $\alpha = 40$, using ours and modified phase-based
    method.
    We also show the magnified result of DMM~\cite{oh2018learning} as a reference result of generic motion magnification.
    Our method generates magnified frames without artifacts and exhibits the $x$-t slice showing clearly legible axial vibrations, while
    modified phase-based
    method
    and DMM both suffer from severe artifacts and have unclear axial vibrations in the $x$-t slice.}
    \label{figure:sep_qual}
\end{figure*}

\paragraph{Qualitative Results}
We demonstrate the advantage of our method that it can amplify only the motion along the axis of interest while disentangling the motions in uninterested
directions that interfere with motion analysis.
To illustrate this concept concretely, consider a scenario where a shaft is rotating in the radial direction.
In such cases, magnifying and examining the motion along the axial direction, which is crucial to assess the condition of the rotating machinery~\cite{luo2021analysis}, becomes challenging due to the dominance of rotational motion over the axial component.
We conduct an experiment shown in Fig.~\ref{figure:sep_qual}
by attaching weights to a rotor to impose an imbalance, which results in axial vibrations.
Then, we acquire a video of the imbalanced rotor, called \textit{rotor imbalance} sequence.
We choose a horizontal-axis line in the original frame and visualize $x$-t slices
for the magnified output frames from each method,
respectively.
Note that we also provide the result of
DMM~\cite{oh2018learning} as a reference to compare the results of axial motion magnification with generic motion magnification.
As shown in Fig.~\ref{figure:sep_qual}, our method produces the magnified output frames without artifacts and exhibits the $x$-t slice that clearly depicts
axial vibrations.
In contrast, the modified phase-based
method suffers from severe ringing artifacts, likely due to the overcompleteness of the complex steerable filter~\cite{simoncelli1992shiftable,simoncelli1995steerable}, which cannot perfectly separate the phase representation into two orthogonal directions.
DMM yields the magnified frames with artifacts and unclear axial vibrations in the $x$-t slice, since the representation of generic motion magnification method struggles to disentangle the dominant motion of the radial direction from the motion of interest, \ie, axial direction's motion.

\paragraph{Quantitative Results}
\begin{figure}[t]
\centering
\includegraphics[width=0.90\linewidth]{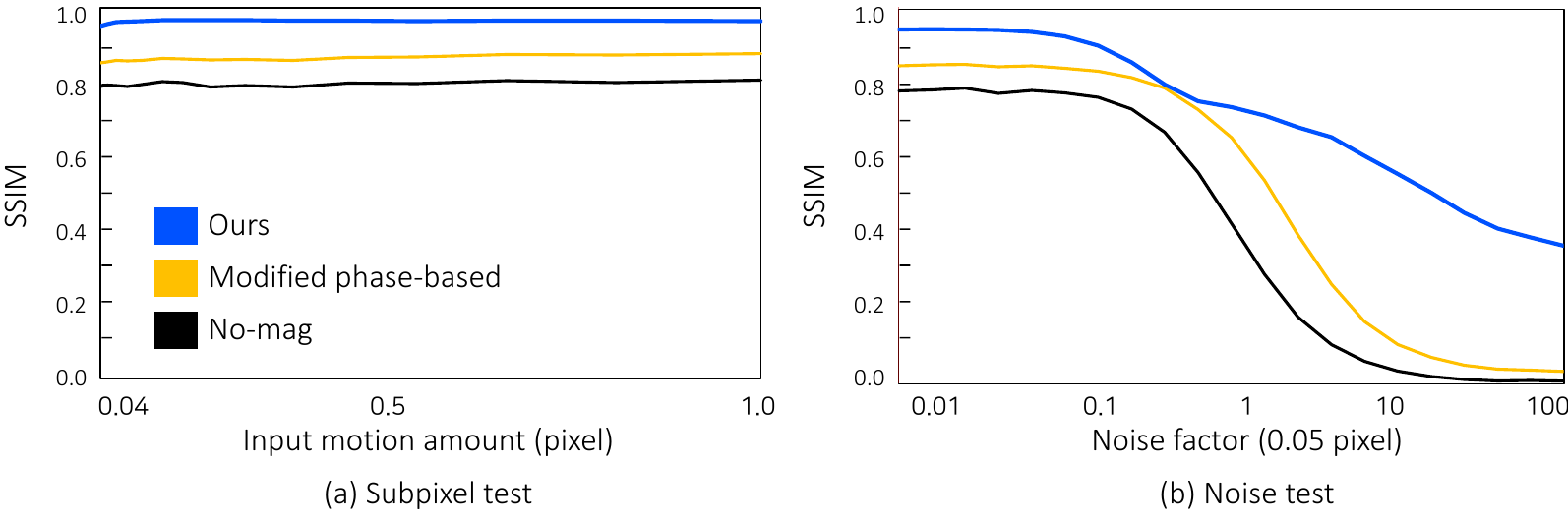}
\caption{
    \textbf{Quantitative results in axial motion magnification scenario.}
    (a) In the subpixel test, ours shows superior performance on SSIM over the modified phase-based
    method across all input motion amount, ranging from $0.04$ to $1.0$.
    (b) In the noise tests when the input motion amount is 0.05 pixel, we observe a growing disparity in SSIM scores between ours and the phase-based approach, as the noise factor rises.
    }
    \label{figure:quantitative2}
\end{figure}

To quantitatively evaluate our learning-based axial motion magnification method, we generate an axial evaluation dataset based on the validation dataset of DMM~\cite{oh2018learning}.
The method of generating the dataset is almost the same as that of the training dataset. One difference is that we adjust the motion amplification factor to ensure that the amplified motion magnitude along a random axis is equal to 10. The motion amplification factor for the other axis is set to half the value. Note that we set $\phi$ to be $0$ for this quantitative evaluation.
We report the Structural Similarity Index (SSIM)~\cite{wang2004image} between the ground truth and output frames of the modified phase-based
method and ours. As a reference, we provide the SSIM between ground truth and input frames.
Figure~\ref{figure:quantitative2} summarizes the results.
We measure the SSIM by varying the levels of motion (Fig.~\ref{figure:quantitative2}-(a) Subpixel test) and additive noise (Fig.~\ref{figure:quantitative2}-(b) Noise test) in the input images. The number of evaluation data samples for each level of motion and noise is $1,000$.
Regardless of the input motion magnitude and noise level, our method consistently outperforms the
modified phase-based
approach, which indicates that our proposed network architecture and dataset are effective for learning axis-wise disentangled representations.

\paragraph{Motion Legibility Comparison}
\begin{figure}[t]
\centering
\includegraphics[width=1\linewidth]{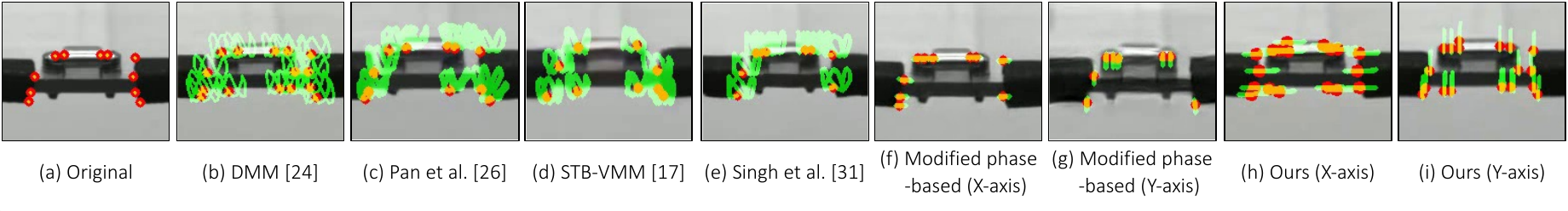}
\caption{
    \textbf{Motion legibility improvement.}
    We visualize the 40$\times$ magnified frames of the structure, which are overlaid with the sampled trajectories from the KLT tracker. Generic motion magnification methods (b-e) produce trajectories that are more complex and hard to interpret. However, Ours (h-i) shows simplified and legible motion trajectories when magnifying along the specific axis (\ie, $x$-axis and $y$-axis), while the modified phase-based methods (f-g)  exhibit small and bounded amplified motions, as theoretically proven in the phase-based method~\cite{wadhwa2013phase}.
    }\label{figure:KLT}
\end{figure}

To demonstrate the improved legibility of magnified motions by our method,
we use a structure that exhibits complex movements.
We then visualize and compare the motion trajectories, tracked by the KLT tracker, of the $40\times$ magnified video sequences of this structure
using both the generic methods and the axial method (Ours). We also provide the modified phase-based for the axial method.
As shown in Fig.~\ref{figure:KLT},
our method shows legible trajectories when magnifying along the specific axis (\ie, $x$-axis and $y$-axis),
while generic motion magnification method (Fig.~\ref{figure:KLT}-(b-e)) shows the entangled trajectories difficult to judge major motion characteristics. The modified phase-based methods (Fig.~\ref{figure:KLT}-(f-g)) result in bounded amplified motion, as theoretically proven in the phase-based method~\cite{wadhwa2013phase}.

\begin{figure}[t]
\centering
\includegraphics[width=1\textwidth]{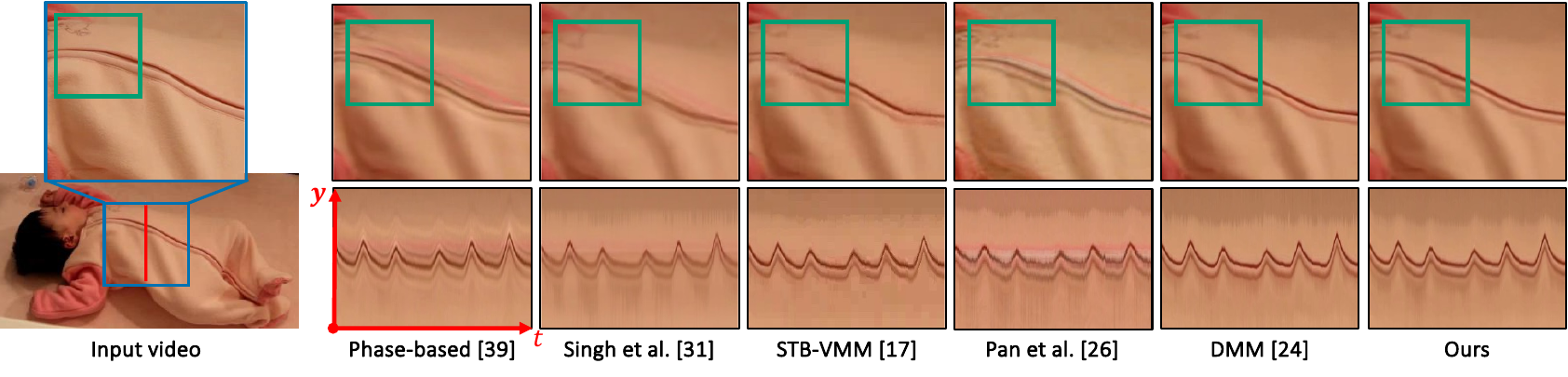}
    \caption{
    \textbf{Qualitative results in generic motion magnification scenario.}
We amplify the \emph{baby} sequence with the magnification factor $\alpha{=}20$, using phase-based method~\cite{wadhwa2013phase}, learning-based methods~\cite{singh2023multi,lado2023stb,oh2018learning,pan2024self}, and Ours.
    Ours and DMM favorably preserve the edges of the clothes and show no ringing artifacts in the magnified frames and the $x$-t slices.
    In contrast, the magnified output frames of the phase-based, Singh~\etal, STB-VMM, and Pan~\etal  show ringing artifacts or blurry results.}
    \label{figure:qual}
\end{figure}

\begin{figure}[t]
\centering
\includegraphics[width=1\linewidth]{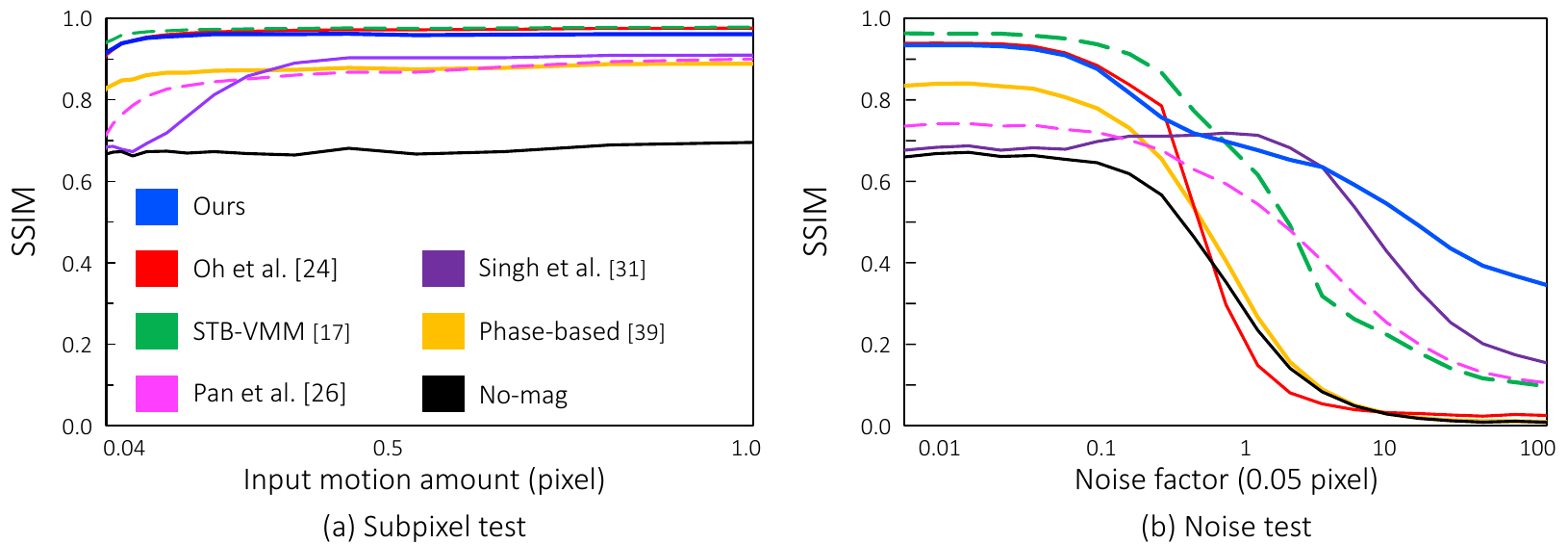}
    \caption{
    \textbf{Quantitative results in generic motion magnification scenario.} (a) In the subpixel test, Ours outperforms phase-based method, Singh~\etal, and Pan~\etal and achieves favorable performance on SSIM compared to DMM and STB-VMM.
    (b) In the noise test, Ours shows comparable noise tolerance compared to other methods and high noise tolerance as the noise factor increases.
    }\label{figure:quantitative1}
\end{figure}

\subsection{Generic Motion Magnification}
Our method can be readily adapted for generic motion magnification scenarios without further training.
This adaptability is achieved by simply multiplying the same magnification factors with the axis-wise shape representations.
In the context of generic motion magnification, we compare our method with the phase-based method~\cite{wadhwa2013phase} and the learning-based methods~\cite{oh2018learning,lado2023stb,singh2023multi,pan2024self}.

\paragraph{Qualitative Results}
We visualize the magnified output frames and plot the $x$-t slices for the \textit{baby} sequence, comparing ours with the several motion magnification methods in the generic
scenarios (see Fig.~\ref{figure:qual}).
Both our method and DMM~\cite{oh2018learning} favorably preserve the edges of the baby's clothing and show no ringing artifacts in the magnified results of breathing motion.
In contrast,
the phase-based method~\cite{wadhwa2013phase}, Singh~\etal~\cite{singh2023multi}, STB-VMM~\cite{lado2023stb}, Pan~\etal~\cite{pan2024self} and show severe ringing artifacts or blurry results\footnote{We reproduced all the results using the codes publicly accessible.}.

\paragraph{Quantitative Results}
To quantitatively verify the ability of our method in generic motion magnification,
we synthesize a generic validation dataset.
Unlike the axial case,
we set the magnification factor $\alpha$ to be identical along the $x$ and $y$ axes.
We report SSIM~\cite{wang2004image} between ground truth and output frames from the phase-based method~\cite{wadhwa2013phase} and the learning-based methods~\cite{oh2018learning,singh2023multi,lado2023stb,pan2024self}.
As shown in Fig.~\ref{figure:quantitative1},
{for}
input motion ranges from $0.04$ to $1.0$, ours outperforms the phase-based method, Singh~\etal~\cite{singh2023multi}, Pan~\etal~\cite{pan2024self}.
Compared to DMM~\cite{oh2018learning} and STB-VMM~\cite{lado2023stb},
ours demonstrates favorable performance,
which exceeds the threshold for visually acceptable SSIM scores~\cite{ha2024revisiting}.
Ours demonstrates comparable noise tolerance to other methods and exhibits high noise tolerance as noise factor increases.

\subsection{Ablation Study}\label{ablation}
In this section, we conduct ablation studies to evaluate the impact of the Motion Separation Module (MSM) and the components of the proposed synthetic training data.
We carry out quantitative experiments on the evaluation dataset of both the generic case and the axial case that has random angles.

\paragraph{Motion Separation Module (MSM)}
\begin{figure}[t]
\centering
\includegraphics[width=0.87\linewidth]{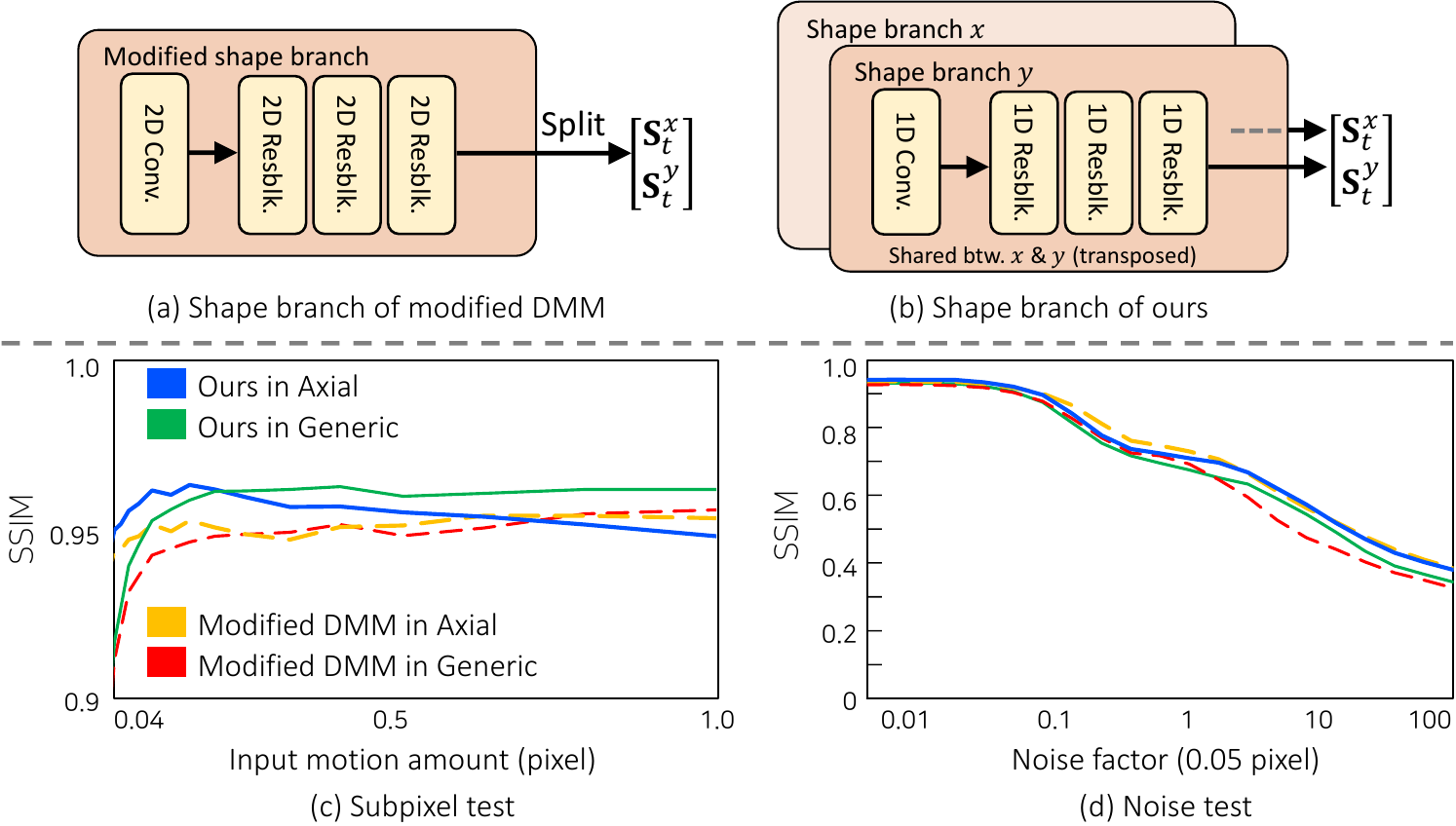}
    \caption{
    \textbf{[Top] Architectural difference on the \emph{shape} branch, [Bottom] Quantitative results of ablating Motion Separation Module (MSM).}
    (a) The modified DMM, designed for ablation study, employs 2D convolutions and splits features along channel dimensions for axial motion magnification.
    (c) Ours with MSM generally achieves higher SSIM in the subpixel test on generic and axial evaluation datasets. (d) In the noise test, Ours shows comparable performance to the modified DMM.
    }
\label{figure:architecture_ablation_seperable}
\end{figure}

To validate the effectiveness of MSM, we design a competitor called modified DMM, which closely resembles that of DMM~\cite{oh2018learning}.
As shown in the top of Fig.~\ref{figure:architecture_ablation_seperable},
different from our method that uses 1D convolutions, the modified DMM employs 2D convolutions in the \emph{Shape} branch and the \emph{Manipulator}.
The axial shape representations of the modified DMM are acquired by dividing the feature map along the channel dimension. We train the networks with the same training details as Ours.
The bottom of Fig.~\ref{figure:architecture_ablation_seperable} shows that
Ours with MSM generally achieves higher SSIM in the generic and axial subpixel tests, which shows the efficacy of the MSM in capturing small motions.
In the noise test, Ours shows comparable performance to the modified DMM.

\begin{figure}[t]
\centering
\includegraphics[width=0.90\linewidth]{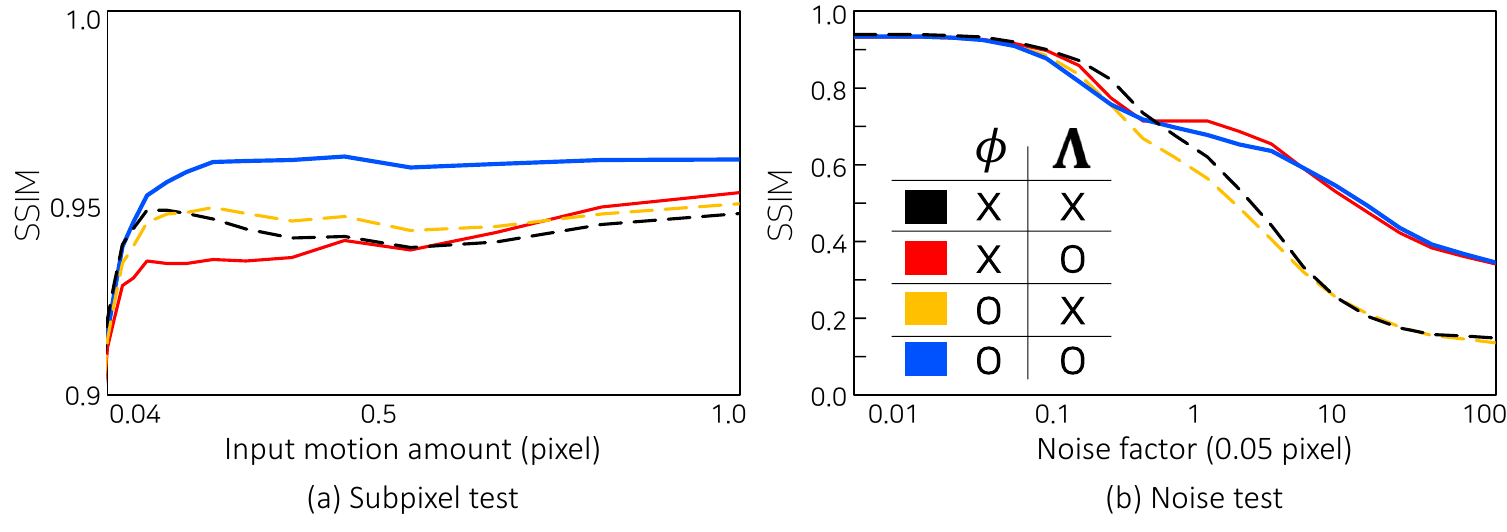}
    \caption{
    \textbf{Ablation study of the components in data generation
    in the generic evaluation dataset.}
    We generate the different
    training data varying the presence of the angle $\phi$ and the object-wise motion magnification map $\boldsymbol{{\Lambda}}$,
    and evaluate
    the networks trained on each dataset configuration using the generic evaluation dataset.
    (a) Using both $\phi$ and $\boldsymbol{{\Lambda}}$ demonstrates best performance in the subpixel test.
    (b) In the noise test, we observe that utilizing $\boldsymbol{{\Lambda}}$ notably enhances noise tolerance.
    }\label{figure:data_ablation}
\end{figure}

\paragraph{Components of Synthetic Training Data}
To evaluate the impact of the angle $\phi$ and the object-wise motion magnification map $\boldsymbol{{\Lambda}}$,
we generate the different types of training data varying the presence of these components.
Our newly designed dataset incorporates both $\phi$ and $\boldsymbol{\Lambda}$, contrasting with the dataset that follows the same setup as DMM~\cite{oh2018learning}, which does not contain either element.
In addition, we generate two more datasets that each add one of these components (\ie, either $\phi$ or $\boldsymbol{\Lambda}$) to the base dataset that initially does not include them.
Note that evaluating the networks trained on these datasets on the axial evaluation dataset is infeasible since the networks trained without $\phi$ cannot perform axial motion magnification. Thus, we use the generic evaluation dataset for this ablation study.
Fig.~\ref{figure:data_ablation} shows that the addition of either $\phi$ or $\boldsymbol{{\Lambda}}$ achieves no
improvement in the subpixel test.
The combined use of both $\phi$ and $\boldsymbol{{\Lambda}}$ yields the most significant performance improvement in the subpixel test, demonstrating that our proposed data set is beneficial in the generic motion magnification task as well.
In the noise test, utilizing $\boldsymbol{{\Lambda}}$  notably enhances noise tolerance, while the addition of $\phi$ has no effect on noise tolerance.

\section{Conclusion}
In this work, we present a novel concept, axial video motion magnification, which improves the legibility of the motions by disentangling and magnifying the motion representations along axes specified by users.
To this end, we propose an innovative learning-based approach for both axial and generic motion magnification, incorporating the Motion Separation Module (MSM) to effectively extract and magnify motion representations along two orthogonal orientations.
To support this, we establish a new synthetic data generation pipeline tailored for the axial motion magnification.
Our method provides user controllability and significantly enhances the legibility of the motions along chosen axes, showing favorable performance compared to competing methods, even in the generic motion magnification case.
Although the axial motion magnification serves as a branch that enhances users' applicability, another branch can be the method to perform motion magnification in real-time, which is useful and beneficial for various applications.
In other respects, most of existing 2D video motion magnification methods including ours assume fixed 2D viewpoints. Exploring real-time inference and moving viewpoints would be a promising direction for future research.

\section*{Acknowledgment}
This work was supported by Institute of Information \& communications Technology Planning \& Evaluation (IITP) grant (No. RS-2022-II220290, Visual Intelligence for Space-Time Understanding and Generation based on Multi-layered Visual Common Sense; No. RS-2019-II191906, Artificial Intelligence Graduate School Program (POSTECH)) and National Research Foundation of Korea (NRF) grant (No. RS-2024-00358135, Corner Vision: Learning to Look Around the Corner through Multi-modal Signals) funded by the Korea government (MSIT).

\bibliographystyle{splncs04}
\bibliography{main}

@STRING{TOG		= "ACM Transactions on Graphics (SIGGRAPH)"}

@STRING{CVPR	= "IEEE Conference on Computer Vision and Pattern Recognition (CVPR)"}

@STRING{ECCV	= "European Conference on Computer Vision (ECCV)"}

@STRING{ICCP	= "IEEE International Conference on Computational Photography (ICCP)"}

@STRING{IS=" IEICE Transactions on Information and Systems. "}

@article{freeman1991motion,
  title={Motion without movement},
  author={Freeman, William T and Adelson, Edward H and Heeger, David J},
  journal={ACM SIGGRAPH},
  volume={25},
  number={4},
  pages={27--30},
  year={1991},
  publisher={ACM New York, NY, USA}
}

@article{li2023computational,
  title={Computational discovery of microstructured composites with optimal stiffness-toughness trade-offs},
  author={Beichen Li and Bolei Deng and Wan Shou and Tae-Hyun Oh and Yuanming Hu and Yiyue Luo and Liang Shi and Wojciech Matusik},
  journal={Science Advances},
  year={2024},
  volume={10},
  number={5}
}

@article{tilbrook2006crack,
  title={Crack propagation paths in layered, graded composites},
  author={Tilbrook, MT and Rozenburg, K and Steffler, ED and Rutgers, L and Hoffman, M},
  journal={Composites Part B: Engineering},
  volume={37},
  number={6},
  pages={490--498},
  year={2006},
  publisher={Elsevier}
}

@article{brodnik2021fracture,
  title={Fracture Diodes: Directional asymmetry of fracture toughness},
  author={Brodnik, NR and Brach, Stella and Long, CM and Ravichandran, G and Bourdin, B and Faber, KT and Bhattacharya, K},
  journal={Physical Review Letters},
  volume={126},
  number={2},
  pages={025503},
  year={2021},
  publisher={APS}
}

@article{wadhwa2013phase,
  title={Phase-based video motion processing},
  author={Wadhwa, Neal and Rubinstein, Michael and Durand, Fr{\'e}do and Freeman, William T},
  journal={ACM Transactions on Graphics (TOG)},
  volume={32},
  number={4},
  pages={1--10},
  year={2013},
  publisher={ACM New York, NY, USA}
}

@inproceedings{wadhwa2014riesz,
  title={Riesz pyramids for fast phase-based video magnification},
  author={Wadhwa, Neal and Rubinstein, Michael and Durand, Fr{\'e}do and Freeman, William T},
  booktitle=ICCP,
  year={2014},
  organization={IEEE}
}

@inproceedings{zhang2017video,
  title={Video acceleration magnification},
  author={Zhang, Yichao and Pintea, Silvia L and Van Gemert, Jan C},
  booktitle=CVPR,
  year={2017}
}

@inproceedings{takeda2018jerk,
  title={Jerk-aware video acceleration magnification},
  author={Takeda, Shoichiro and Okami, Kazuki and Mikami, Dan and Isogai, Megumi and Kimata, Hideaki},
  booktitle=CVPR,
  year={2018}
}

@inproceedings{takeda2019video,
  title={Video magnification in the wild using fractional anisotropy in temporal distribution},
  author={Takeda, Shoichiro and Akagi, Yasunori and Okami, Kazuki and Isogai, Megumi and Kimata, Hideaki},
  booktitle=CVPR,
  year={2019}
}

@article{takeda2020local,
  title={Local Riesz pyramid for faster phase-based video magnification},
  author={Takeda, Shoichiro and Isogai, Megumi and Shimizu, Shinya and Kimata, Hideaki},
  journal=IS,
  volume={103},
  number={10},
  pages={2036--2046},
  year={2020},
  publisher={The Institute of Electronics, Information and Communication Engineers}
}

@inproceedings{takeda2022bilateral,
  title={Bilateral Video Magnification Filter},
  author={Takeda, Shoichiro and Niwa, Kenta and Isogawa, Mariko and Shimizu, Shinya and Okami, Kazuki and Aono, Yushi},
  booktitle=CVPR,
  year={2022}
}

@article{wu2012eulerian,
  title={Eulerian video magnification for revealing subtle changes in the world},
  author={Wu, Hao-Yu and Rubinstein, Michael and Shih, Eugene and Guttag, John and Durand, Fr{\'e}do and Freeman, William},
  journal={ACM transactions on graphics (TOG)},
  volume={31},
  number={4},
  pages={1--8},
  year={2012},
  publisher={ACM New York, NY, USA}
}

@inproceedings{oh2018learning,
  title={Learning-based video motion magnification},
  author={Oh, Tae-Hyun and Jaroensri, Ronnachai and Kim, Changil and Elgharib, Mohamed and Durand, Fr'edo and Freeman, William T and Matusik, Wojciech},
  booktitle={Proceedings of the European Conference on Computer Vision (ECCV)},
  pages={633--648},
  year={2018}
}

@InProceedings{COCO,
  title={Microsoft coco: Common objects in context},
  author={Lin, Tsung-Yi and Maire, Michael and Belongie, Serge and Hays, James and Perona, Pietro and Ramanan, Deva and Doll{\'a}r, Piotr and Zitnick, C Lawrence},
  booktitle=ECCV,
  year={2014}
}

@article{PASCAL,
  title={The pascal visual object classes (voc) challenge},
  author={Everingham, Mark and Van Gool, Luc and Williams, Christopher KI and Winn, John and Zisserman, Andrew},
  journal={International journal of computer vision},
  volume={88},
  number={2},
  pages={303--338},
  year={2010}
}

@article{liu2005motion,
  title={Motion magnification},
  author={Liu, Ce and Torralba, Antonio and Freeman, William T and Durand, Fr{\'e}do and Adelson, Edward H},
  journal={ACM transactions on graphics (TOG)},
  volume={24},
  number={3},
  pages={519--526},
  year={2005},
  publisher={ACM New York, NY, USA}
}

@article{simoncelli1992shiftable,
  title={Shiftable multiscale transforms},
  author={Simoncelli, Eero P and Freeman, William T and Adelson, Edward H and Heeger, David J},
  journal={IEEE transactions on Information Theory},
  volume={38},
  number={2},
  pages={587--607},
  year={1992},
  publisher={IEEE}
}

@inproceedings{simoncelli1995steerable,
  title={The steerable pyramid: A flexible architecture for multi-scale derivative computation},
  author={Simoncelli, Eero P and Freeman, William T},
  booktitle={Proceedings., International Conference on Image Processing},
  volume={3},
  pages={444--447},
  year={1995},
  organization={IEEE}
}

@article{cha2017output,
  title={Output-only computer vision based damage detection using phase-based optical flow and unscented Kalman filters},
  author={Cha, Y-J and Chen, Justin G and B{\"u}y{\"u}k{\"o}zt{\"u}rk, Oral},
  journal={Engineering Structures},
  volume={132},
  pages={300--313},
  year={2017},
  publisher={Elsevier}
}

@inproceedings{chen2014structural,
  title={Structural modal identification through high speed camera video: Motion magnification},
  author={Chen, Justin G and Wadhwa, Neal and Cha, Young-Jin and Durand, Fr{\'e}do and Freeman, William T and Buyukozturk, Oral},
  booktitle={Topics in Modal Analysis I, Volume 7: Proceedings of the 32nd IMAC, A Conference and Exposition on Structural Dynamics, 2014},
  pages={191--197},
  year={2014},
  organization={Springer}
}

@inproceedings{balakrishnan2013detecting,
  title={Detecting pulse from head motions in video},
  author={Balakrishnan, Guha and Durand, Fredo and Guttag, John},
  booktitle={Proceedings of the IEEE conference on computer vision and pattern recognition},
  pages={3430--3437},
  year={2013}
}

@article{vernekar2014gear,
  title={Gear fault detection using vibration analysis and continuous wavelet transform},
  author={Vernekar, Kiran and Kumar, Hemantha and Gangadharan, KV},
  journal={Procedia Materials Science},
  volume={5},
  pages={1846--1852},
  year={2014},
  publisher={Elsevier}
}

@article{davis2014visual,
  title={The visual microphone: Passive recovery of sound from video},
  author={Davis, Abe and Rubinstein, Michael and Wadhwa, Neal and Mysore, Gautham J and Durand, Fredo and Freeman, William T},
  journal=TOG,
  year={2014},
  publisher={Association for Computing Machinery (ACM)}
}

@article{freeman1991design,
  title={The design and use of steerable filters},
  author={Freeman, William T and Adelson, Edward H and others},
  journal={IEEE Transactions on Pattern analysis and machine intelligence},
  volume={13},
  number={9},
  pages={891--906},
  year={1991}
}

@article{lado2023stb,
  title={STB-VMM: Swin Transformer based Video Motion Magnification},
  author={Lado-Roig{\'e}, Ricard and P{\'e}rez, Marco A},
  journal={Knowledge-Based Systems},
  volume={269},
  pages={110493},
  year={2023},
  publisher={Elsevier}
}

@inproceedings{singh2023multi,
  title={Multi Domain Learning for Motion Magnification},
  author={Singh, Jasdeep and Murala, Subrahmanyam and Kosuru, G},
  booktitle={Proceedings of the IEEE/CVF Conference on Computer Vision and Pattern Recognition},
  pages={13914--13923},
  year={2023}
}

@article{oliveto1997complex,
  title={Complex modal analysis of a flexural vibrating beam with viscous end conditions},
  author={Oliveto, G and Santini, Adolfo and Tripodi, E},
  journal={Journal of Sound and Vibration},
  volume={200},
  number={3},
  pages={327--345},
  year={1997},
  publisher={Elsevier}
}

@inproceedings{lucas1981iterative,
  title={An iterative image registration technique with an application to stereo vision},
  author={Lucas, Bruce D and Kanade, Takeo},
  booktitle={IJCAI'81: 7th international joint conference on Artificial intelligence},
  volume={2},
  pages={674--679},
  year={1981}
}

@article{luo2021analysis,
  title={Analysis of vibration characteristics of centrifugal pump mechanical seal under wear and damage degree},
  author={Luo, Yin and Zhang, Wenqi and Fan, Yakun and Han, Yuejiang and Li, Weimin and Acheaw, Emmanuel},
  journal={Shock and Vibration},
  volume={2021},
  pages={1--9},
  year={2021},
  publisher={Hindawi Limited}
}

@article{chen2017video,
  title={Video camera--based vibration measurement for civil infrastructure applications},
  author={Chen, Justin G and Davis, Abe and Wadhwa, Neal and Durand, Fr{\'e}do and Freeman, William T and B{\"u}y{\"u}k{\"o}zt{\"u}rk, Oral},
  journal={Journal of Infrastructure Systems},
  volume={23},
  number={3},
  pages={B4016013},
  year={2017},
  publisher={American Society of Civil Engineers}
}

@incollection{chen2015developments,
  title={Developments with motion magnification for structural modal identification through camera video},
  author={Chen, Justin G and Wadhwa, Neal and Durand, Fr{\'e}do and Freeman, William T and Buyukozturk, Oral},
  booktitle={Dynamics of Civil Structures, Volume 2},
  pages={49--57},
  year={2015},
  publisher={Springer}
}

@article{chen2015modal,
  title={Modal identification of simple structures with high-speed video using motion magnification},
  author={Chen, Justin G and Wadhwa, Neal and Cha, Young-Jin and Durand, Fr{\'e}do and Freeman, William T and Buyukozturk, Oral},
  journal={Journal of Sound and Vibration},
  volume={345},
  pages={58--71},
  year={2015},
  publisher={Elsevier}
}

@article{brattoli2021unsupervised,
  title={Unsupervised behaviour analysis and magnification (uBAM) using deep learning},
  author={Brattoli, Biagio and B{\"u}chler, Uta and Dorkenwald, Michael and Reiser, Philipp and Filli, Linard and Helmchen, Fritjof and Wahl, Anna-Sophia and Ommer, Bj{\"o}rn},
  journal={Nature Machine Intelligence},
  volume={3},
  number={6},
  pages={495--506},
  year={2021},
  publisher={Nature Publishing Group UK London}
}

@article{moya2020non,
  title={A non-contact heart rate estimation method using video magnification and neural networks},
  author={Moya-Albor, Ernesto and Brieva, Jorge and Ponce, Hiram and Mart{\'\i}nez-Villase{\~n}or, Lourdes},
  journal={IEEE Instrumentation \& Measurement Magazine},
  volume={23},
  number={4},
  pages={56--62},
  year={2020},
  publisher={IEEE}
}

@article{fan2021robotically,
  title={Robotically surgical vessel localization using robust hybrid video motion magnification},
  author={Fan, Wenkang and Zheng, Zhuohui and Zeng, Wankang and Chen, Yinran and Zeng, Hui-Qing and Shi, Hong and Luo, Xiongbiao},
  journal={IEEE Robotics and Automation Letters},
  volume={6},
  number={2},
  pages={1567--1573},
  year={2021},
  publisher={IEEE}
}

@inproceedings{janatka2020surgical,
  title={Surgical video motion magnification with suppression of instrument artefacts},
  author={Janatka, Mirek and Marcus, Hani J and Dorward, Neil L and Stoyanov, Danail},
  booktitle={Medical Image Computing and Computer Assisted Intervention--MICCAI 2020: 23rd International Conference, Lima, Peru, October 4--8, 2020, Proceedings, Part III 23},
  pages={353--363},
  year={2020},
  organization={Springer}
}

@article{qiu2018defect,
  title={Defect detection in FRP-bonded structural system via phase-based motion magnification technique},
  author={Qiu, Qiwen and Lau, Denvid},
  journal={Structural Control and Health Monitoring},
  volume={25},
  number={12},
  pages={e2259},
  year={2018},
  publisher={Wiley Online Library}
}

@article{smieja2021motion,
  title={Motion Magnification of Vibration Image in Estimation of Technical Object Condition-Review},
  author={{\'S}mieja, Micha{\l} and Mamala, Jaros{\l}aw and Pra{\.z}nowski, Krzysztof and Ciepli{\'n}ski, Tomasz and Szumilas, {\L}ukasz},
  journal={Sensors},
  volume={21},
  number={19},
  pages={6572},
  year={2021},
  publisher={MDPI}
}

@article{sarrafi2018vibration,
  title={Vibration-based damage detection in wind turbine blades using Phase-based Motion Estimation and motion magnification},
  author={Sarrafi, Aral and Mao, Zhu and Niezrecki, Christopher and Poozesh, Peyman},
  journal={Journal of Sound and vibration},
  volume={421},
  pages={300--318},
  year={2018},
  publisher={Elsevier}
}

@inproceedings{ghiasi2021simple,
  title={Simple copy-paste is a strong data augmentation method for instance segmentation},
  author={Ghiasi, Golnaz and Cui, Yin and Srinivas, Aravind and Qian, Rui and Lin, Tsung-Yi and Cubuk, Ekin D and Le, Quoc V and Zoph, Barret},
  booktitle={Proceedings of the IEEE/CVF conference on computer vision and pattern recognition},
  pages={2918--2928},
  year={2021}
}

@article{wang2004image,
  title={Image quality assessment: from error visibility to structural similarity},
  author={Wang, Zhou and Bovik, Alan C and Sheikh, Hamid R and Simoncelli, Eero P},
  journal={IEEE transactions on image processing},
  volume={13},
  number={4},
  pages={600--612},
  year={2004},
  publisher={IEEE}
}

@article{pan2024self,
  title={Self-Supervised Motion Magnification by Backpropagating Through Optical Flow},
  author={Pan, Zhaoying and Geng, Daniel and Owens, Andrew},
  journal={Advances in Neural Information Processing Systems},
  volume={36},
  year={2024}
}

@misc{ha2024revisiting,
      title={Revisiting Learning-based Video Motion Magnification for Real-time Processing},
      author={Hyunwoo Ha and Oh Hyun-Bin and Kim Jun-Seong and Kwon Byung-Ki and Kim Sung-Bin and Linh-Tam Tran and Ji-Yun Kim and Sung-Ho Bae and Tae-Hyun Oh},
      year={2024},
      eprint={2403.01898},
      archivePrefix={arXiv},
      primaryClass={cs.CV}
}

@article{yang2005vibex,
  title={VIBEX: an expert system for vibration fault diagnosis of rotating machinery using decision tree and decision table},
  author={Yang, Bo-Suk and Lim, Dong-Soo and Tan, Andy Chit Chiow},
  journal={Expert Systems with Applications},
  volume={28},
  number={4},
  pages={735--742},
  year={2005},
  publisher={Elsevier}
}
\clearpage
\begin{center}
{\Large\bfseries Learning-based Axial Video Motion Magnification\par}
\vspace{0.6em}
{\large\bfseries Supplementary Material\par}
\end{center}
\vspace{0.6em}

\section*{Contents}

\noindent\hyperref[sec:A]{\textbf{A \ \ \ Implementation Details}}\\
\hyperref[sec:A.1]{\text{\qquad A.1 \ \ \ Data Generation}} \\
\hyperref[sec:A.2]{\text{\qquad A.2 \ \ \ Loss Function}} \\
\hyperref[sec:A.3]{\text{\qquad A.3 \ \ \ Projection Layer}} \\
\hyperref[sec:B]{\textbf{B \ \ \ Additional Experiments}} \\
\hyperref[sec:B.1]{\text{\qquad B.1 \ \ \
Quantitative Results Considering Practical Use
}} \\
\hyperref[sec:B.2]{\text{\qquad B.2 \ \ \
Quantitative Results on a Semi-realistic Dataset
}} \\
\hyperref[sec:B.3]{\text{\qquad B.3 \ \ \
Physical Accuracy of Generic Motion Magnification}} \\
\hyperref[sec:B.4]{\text{\qquad B.4 \ \ \
Physical Accuracy of Axial Motion Magnification
}} \\
\hyperref[sec:B.5]{\text{\qquad B.5 \ \ \
Analysis of Amplifiable Motion Magnitudes
}} \\
\hyperref[sec:B.6]{\text{\qquad B.6 \ \ \
Motion Separation Effect of the MSM
}} \\
\hyperref[sec:B.7]{\text{\qquad B.7 \ \ \ Angular Analysis of Axial Motion Magnification
}} \\
\hyperref[sec:B.8]{\text{\qquad B.8 \ \ \ Per-pixel Motion Magnification
}} \\
\hyperref[sec:B.9]{\text{\qquad B.9 \ \ \ Comparison of Inference Speed
}} \\
\hyperref[sec:C]{\textbf{C \ \ \ Additional Results on Diverse Scenarios}} \\

\noindent\rule{\linewidth}{0.2pt}
\section*{Supplementary Material}
In this supplementary material, we present the implementation
details
and additional experiments.
Furthermore, we provide axial and generic motion magnification results across various scenarios.
\appendix
\renewcommand{\theHsection}{appendix.\Alph{section}}

\section{Implementation Details}\label{sec:A}
We provide the details of data generation pipeline (Sec.~\ref{sec:A.1}), the loss function for the learning-based axial motion magnification (Sec.~\ref{sec:A.2}) and the details of projection layer (Sec.~\ref{sec:A.3}).

\subsection{Data generation}\label{sec:A.1}

\paragraph{Training Dataset}
We randomly sample
foreground textures
{ranging from $7$ to $14$}
with segmentation masks from PASCAL VOC~\cite{PASCAL} and one background from COCO~\cite{COCO}. For each layer, we sample the axial magnification factor $\boldsymbol{\alpha}=(\alpha^{\phi};\alpha^{\phi_\perp})$ from the uniform distribution
{whose values are}
ranging from $1$ to $80$.
Each element of translation parameter $\textbf{d}\in \Real^2$ is uniformly sampled from the range $-u$ to $u$, where $u = \mbox{min}(10,30/\mbox{max}(\alpha^{\phi},\alpha^{\phi_\perp}))$.
It limits input motions to a maximum of $10$ pixels or ensures amplified motions are kept under $30$ pixels.
We sample the angle $\phi$ within the range of $0$ to $90$ degrees.
Note that our method enables axial motion magnification not only in the angle $\phi$ but also in the angle $\phi_\perp$,
thus facilitating axial motion magnification within the range of $0$ to $180$ degrees.
To address the loss of subpixel motion due to image quantization, as proposed in DMM~\cite{oh2018learning}, we apply uniform quantization noise to the images before quantizing them.

\paragraph{Generic Evaluation Dataset}
Based on the validation dataset of DMM~\cite{oh2018learning}, we construct the generic evaluation dataset comprising the previous image, next image, magnified image, and a single magnification factor. The generic evaluation dataset consists of two datasets for the subpixel test and noise test.
The dataset for the subpixel test includes $15$ levels of motion, ranging from a motion magnitude of $0.04$ to $1.0$ pixel, changing in a logarithmic scale. The motion magnification factor is adjusted to ensure that the amplified motion magnitude becomes $10$ pixel.
The dataset for the noise test includes $21$ levels of noise, ranging from a noise factor of $0.01$ to $100$ in a logarithmic scale. The amount of input motion is $0.05$ pixel, and the motion amplification factor is also set to ensure that the amplified motion magnitude becomes $10$ pixel.

\paragraph{Axial Evaluation Dataset}
The axial evaluation dataset consists of the previous image, next image, axially magnified image, axial magnification factor vector, and angle.
The axial magnification factor vector is composed of two magnification factors corresponding to two orthogonal orientations.
The axial evaluation dataset also includes two datasets for the subpixel test and noise test.
For the subpixel test dataset, we generate data with 15 levels of motion ranging from $0.04$ to $1.0$ pixel in a logarithmic scale.
We set the motion amplification factor vector to guarantee that the magnified motion magnitude along a random orientation equals 10 pixel. For the other orientation axis, we allocate half of that value.
For the noise test dataset, we have $21$ levels of noise factor ranging from $0.01$ to $100$ in a logarithmic scale. The input motion size along two orthogonal orientations is 0.05 pixel, and the motion magnification factor is set to achieve an amplified motion size of $10$ pixel for one of the orthogonal axes, while the motion magnification factor for the other axis is set to half of that value.
The angle $\phi$ is randomly sampled between $0$ and $90$ degrees, except in the experiment of Fig.~5 in the main paper, where $\phi$ is set to the $0$ degrees for comparison with the phase-based method~\cite{wadhwa2013phase}.

\subsection{Loss Function}\label{sec:A.2}
DMM~\cite{oh2018learning} proposes the texture loss $L_{\text{texture}}$ and shape loss $L_{\text{shape}}$ to represent intensity and motion information, respectively.
These losses are combined with the reconstruction loss $L_{\text{recon}}$,
forming the composite loss function of DMM.
We slightly modify the loss of DMM to separately impose the loss to the x-axis and y-axis shape representations.
The total loss function $L_{\text{total}}$ is as follows:
\begin{align}
\hspace{-4mm}
    L_\text{total} &= L_\text{recon}(\hat{\rmI}^{\phi},\tilde{\rmI}^{\phi}) + \beta(L_\text{texture}(\textbf{T}_{1},\textbf{T}_{2})
    + L_\text{shape}(\textbf{S}^x_{2},\acute{\textbf{S}}^x_{2})) + L_\text{shape}(\textbf{S}^y_{2},\acute{\textbf{S}}^y_{2}),
\end{align}
where we set $\beta$ to 0.5. We train our model using two NVIDIA Titan RTX GPUs.

\subsection{Projection Layer}\label{sec:A.3}
\begin{figure}[t]
\includegraphics[width=1\linewidth]{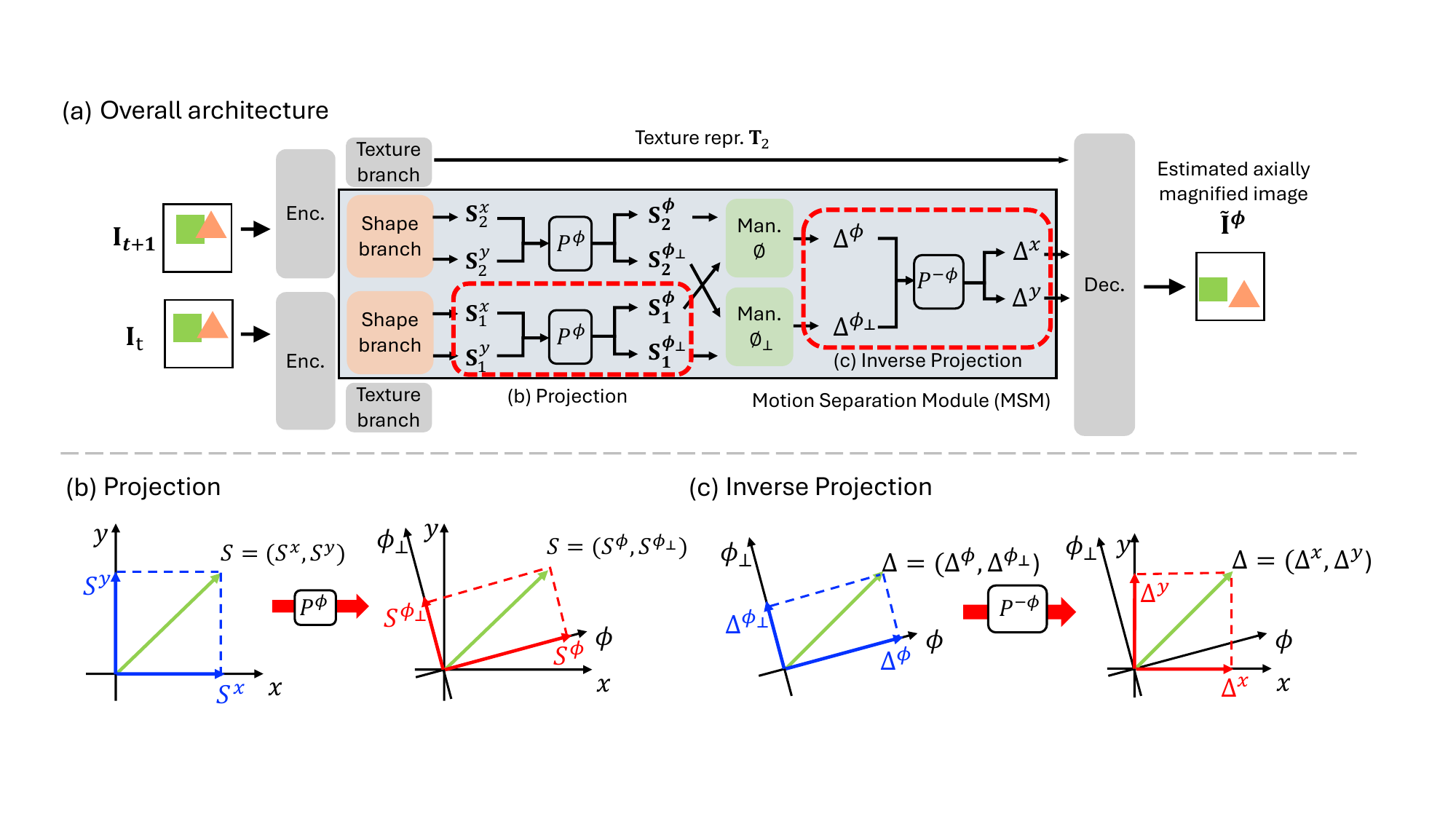}
    \caption{
    \textbf{Projection layer.}
The projection and inverse projection layers facilitate the synthesis of arbitrarily rotated representations through a linear combination. In (a), the representations aligned with the $x$ and $y$-axes undergo projection onto the $\phi$ and $\phi_\perp$ directions. Within the $\phi$ and $\phi_\perp$ directions, these representations are manipulated before being projected back onto the $x$ and $y$-axes, \ie, inverse projection (c).
    }\label{figure:projection}
\end{figure}

Motivated by the concept of steerable filters~\cite{freeman1991design}, we design the projection layer $P^{\phi}$ and inverse projection layer $P^{-\phi}$
using
linear matrices.
It
enables the synthesis of arbitrarily rotated representations through a linear combination of directional representations.
As shown in Fig.~\ref{figure:projection}-(a), the axial shape representations along the canonical $x$ and $y$-axes, which is induced by weight-shared 1D convolutions, are fed to the projection layer $P^{\phi}$. With the linear operation, $P^{\phi}$ projects them and results in the axial shape representations of $\phi$ and $\phi_\perp$ directions (Fig.~\ref{figure:projection}-(b)).
Conversely, the inverse projection layer $P^{-\phi}$ projects the outputs of the \textit{Manipulator} $\Delta^\phi,\Delta^{\phi_\perp}$ back to the canonical $x$ and $y$-axes (Fig.~\ref{figure:projection}-(c)).

\section{Additional Experiments}\label{sec:B}
In this section, we provide two additional quantitative experiments: one assuming the practical use of magnification factors (Sec.~\ref{sec:B.1}) and the other using a semi-realistic dataset rendered by Blender (Sec.~\ref{sec:B.2}). We also assess the physical accuracy of generic motion magnification (Sec.~\ref{sec:B.3}),
the physical accuracy of the proposed axial motion magnification (Sec.~\ref{sec:B.4}), amplifiable motion magnitudes of magnification methods (Sec.~\ref{sec:B.5}),
motion separation effect of the MSM (Sec.~\ref{sec:B.6}), and the behavior of our method across varying degrees (Sec.~\ref{sec:B.7}). We also demonstrate the per-pixel motion magnification capability of our method (Sec.~\ref{sec:B.8}). Finally, we report the inference speed of our method and learning-based motion magnification methods~\cite{oh2018learning,lado2023stb,singh2023multi,pan2024self} (Sec.~\ref{sec:B.9}).

\begin{figure}[t]
\includegraphics[width=1\linewidth]{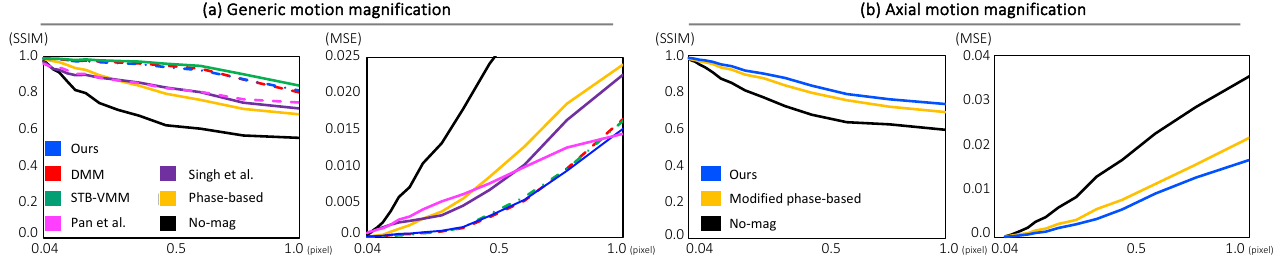}
\centering
    \caption{
    \textbf{Quantitative results in the practical magnification factors.} (a) In this generic subpixel test, Ours achieves higher performance compared to the phase-based method~\cite{wadhwa2013phase}, Singh~\etal~\cite{singh2023multi}, and Pan~\etal~\cite{pan2024self}, and shows comparable performance compared to DMM~\cite{oh2018learning} and STB-VMM~\cite{lado2023stb}. (b) In the axial subpixel test, Ours outperforms the modified phase-based method.
    }
    \label{figure:practical_use}
\end{figure}

\begin{figure}[t]
\includegraphics[width=0.7\linewidth]{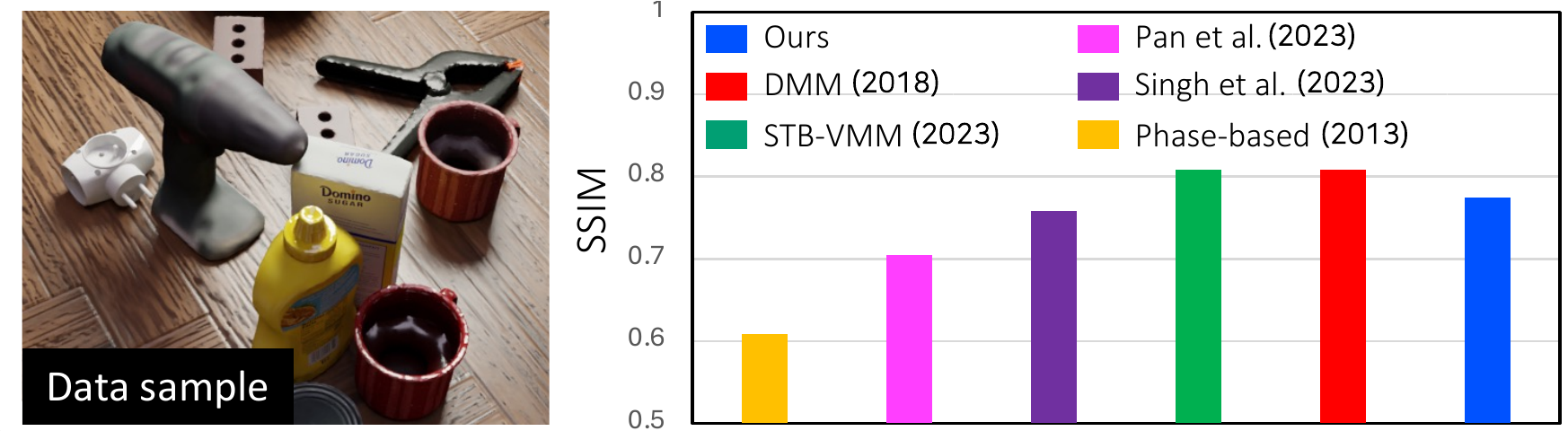}
\centering
    \caption{
    \textbf{Quantitative results on a semi-realistic dataset.
    } With Blender, we generate the semi-realistic evaluation dataset and evaluate the generic motion magnification performance. The results show the consistent tendency of the generic subpixel test using synthetic datasets (Fig.~8 of the main paper). Ours achieves favorable third-ranked performance. However, overall performance drops compared to the synthetic dataset evaluations due to 3D motion and realistic effects.
    }\label{figure:semireal}
\end{figure}

\subsection{Quantitative Results Considering Practical Use}\label{sec:B.1}
In the quantitative results in Fig.~8 of the main paper, we strictly follow the evaluation setting proposed by DMM~\cite{oh2018learning}. This setting is designed to analyze model performance up to its limit. However, such a large magnification factor (e.g., 250) would rarely be used in many applications. Thus, we generate new generic and axial subpixel evaluation datasets with magnification factors arbitrarily ranging from 10 to 50, regardless of input motion. In these subpixel experiments, we use the SSIM and mean squared error (MSE) as evaluation metrics. As shown in Fig.~\ref{figure:practical_use}, our method achieves favorable performance compared to the competing methods with the practical magnification factors.

\subsection{Quantitative Results on a Semi-realistic Dataset}\label{sec:B.2}
Because the training dataset of DMM~\cite{oh2018learning} is publicly available, most learning-based motion magnification methods, such as DMM, STB-VMM~\cite{lado2023stb}, and Singh~\etal~\cite{singh2023multi}, follow the same training procedure. However, since the DMM evaluation dataset has not been published, each method evaluates its performance using its own criteria. To ensure fair comparisons with other methods trained on the DMM training dataset, we strictly follow the DMM data generation pipeline and propose our training and evaluation datasets.
Although Pan~\etal~\cite{pan2024self} evaluated their method on DMM’s data, their model was trained using unlabeled video.
For a further fair comparison, we newly build a semi-realistic dataset for the generic subpixel test using Blender, by randomly moving 3D objects within an input motion range of 1.5 pixels and with random magnification factors between 10 and 30.
The dataset contains 200 samples. As shown in Fig.~\ref{figure:semireal}, our method achieves favorable third-ranked performance, consistent with the generic subpixel test using synthetic datasets (Fig.~8 of the main paper). However, all methods show relatively lower SSIMs compared to the subpixel test using synthetic datasets due to 3D motion and realistic effects.

\begin{figure*}[t]
\includegraphics[width=1\textwidth]{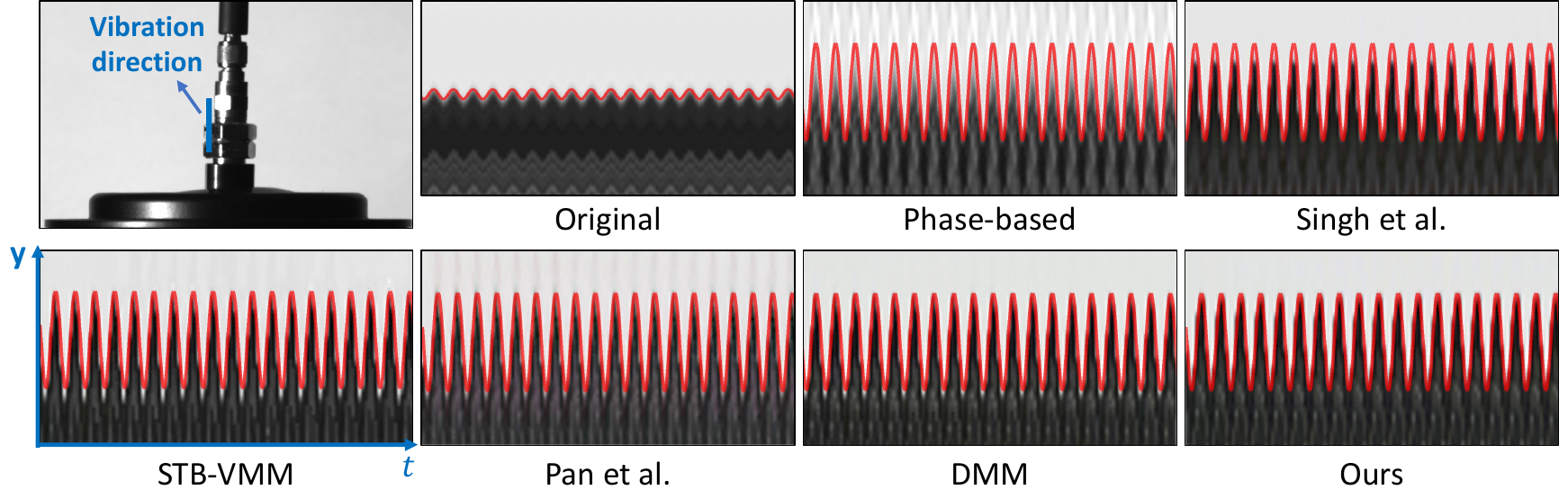}
\centering
    \caption{
    \textbf{Physical accuracy on generic motion magnification.} We compare the physically calculated sinusoidal wave of pixel displacement (red line) to the $y$-t slice's waves of $10\times$ magnified videos from motion magnification methods. We also provide the $y$-t slice's wave of the original video and sinusoidal wave before amplification
    for reference.
    The $y$-t slice's wave of Ours matches the actual pixel displacement.
    The phase-based method~\cite{wadhwa2013phase} exhibits results consistent with the red wave of pixel displacement, albeit suffering from ringing artifacts. Other learning-based methods, such as DMM~\cite{oh2018learning}, STB-VMM~\cite{lado2023stb}, Pan~\etal~\cite{pan2024self}, and Singh~\etal\cite{singh2023multi}, also demonstrate correspondences, with a marginal difference in amplification.
    }
    \label{figure:physical}
\end{figure*}

\begin{table}[t]
\centering
\caption{\textbf{Hyperparameters for acquiring pixel displacement.}}
\begin{tabular}{ccc}
    \toprule
    Hyperparameters & Unit & Value\\
    \midrule
    Vibration frequency $\omega$ & Hz & 20  \\
    Peak amplitude of acceleration $a$ & $\text{m/s}^2$ & 4.11  \\
    Camera-to-vibrator distance $L$ & $\text{m}$ & 2  \\
    Focal length $f$ & $\text{mm}$ & 100  \\
    Per-pixel sensor size $v$ & $\mu$$\text{m}$ & 5.86  \\
    \bottomrule
\end{tabular}
\label{tab:physical_tab}
\end{table}

\subsection{Physical Accuracy of Generic Motion Magnification}\label{sec:B.3}
To assess the physical accuracy of each method on the generic motion magnification scenario, we examine whether the vibrations of the video, which are magnified by each method, match those of actual vibrations.
First, we generate a $20$Hz sinusoidal vibration using a vibration generator. Next, we obtain the peak amplitude of acceleration ($\text{m}/\text{s}^2$) from the attached accelerometer and convert it into a sinusoidal wave of displacement ($\text{m}$), which is transformed into a sinusoidal wave of pixel displacement ($\text{px}$) on the image plane through pinhole camera geometry.
We investigate whether this wave corresponds to the vibration of the $10\times$ magnified video using the \textit{static} mode. The transformation from the peak amplitude of acceleration $a$ to the peak amplitude of displacement $\mu$ is as follows:
\begin{equation}
    \mu = a / \omega^2,
\end{equation}
where $\omega$ denotes the frequency of sinusoidal vibration.
Using $\mu$, we obtain the sinusoidal wave of real-world displacement $s(t)$ over time $t$ and transform it into
pixel displacement $k(t)$, which corresponds to
\begin{equation}
    k(t) = \frac{f}{Lv}s(t).
\end{equation}
The $f$, $L$, and $v$ refer to the focal length, camera-to-vibrator distance, and per-pixel sensor size.

As shown in Fig.~\ref{figure:physical}, the
sinusoidal wave of
our method demonstrates a correspondence with the red wave of pixel displacement that is $10\times$ amplified.
The phase-based method~\cite{wadhwa2013phase} and other learning-based methods~\cite{oh2018learning,singh2023multi,lado2023stb,pan2024self} also exhibit correspondences, albeit with slight differences in amplification.
These results validate the physical accuracy of our method, as well as that of other motion amplification methods.
We provide the hyperparameters for converting acceleration ($\text{m}/\text{s}^{2}$) to pixel displacement (px) in Table~\ref{tab:physical_tab}.

\begin{figure}[t]
\includegraphics[width=1\linewidth]{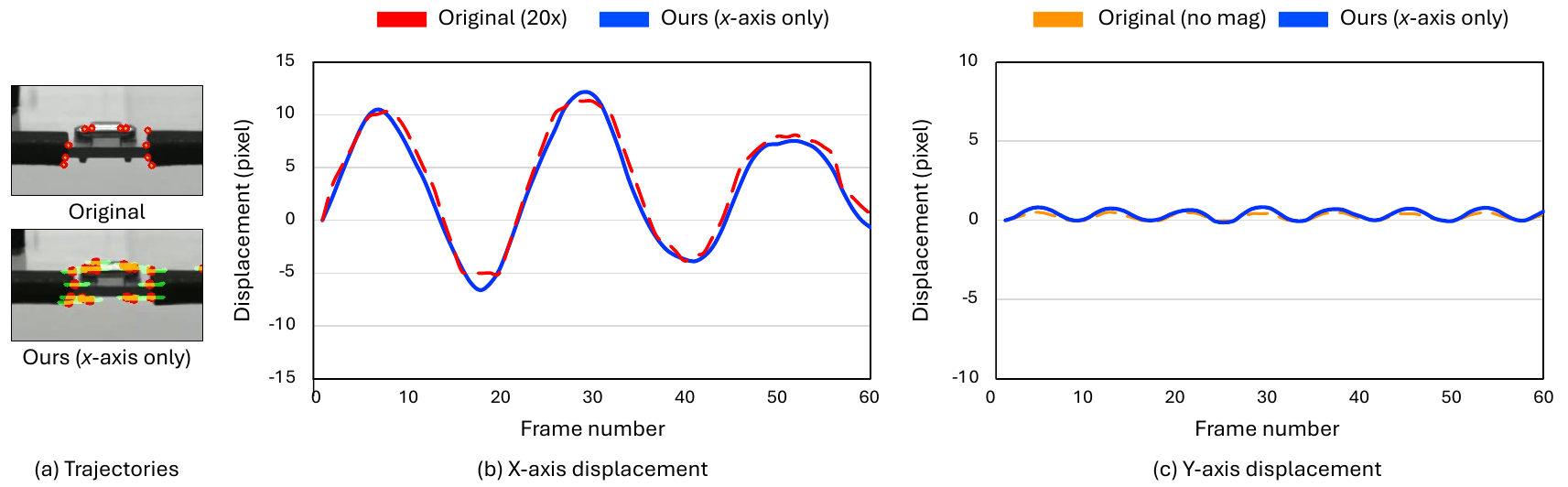}
\centering
    \caption{
    \textbf{Physical accuracy of the proposed axial motion magnification.}
    (a) Using the Kanade-Lucas-Tomasi (KLT) Tracker~\cite{lucas1981iterative}, we obtain the displacement values of the original video and the video that is 20$\times$ amplified along $x$-axis by our method.
    (b) We multiply the $x$-axis' displacement value of the original trajectory by $20$ and compare it with the $x$-axis' displacement value of the video that is amplified along x-axis by our method.
    (c) In the $y$-axis direction, we compare the $y$-axis' displacement of the original trajectory by $20$ with the $y$-axis' displacement value of the video which is amplified along $x$-axis by our method.
    }
    \label{figure:klt_analysis}
\end{figure}

\subsection{Physical Accuracy of Axial Motion Magnification}\label{sec:B.4}
We assess the physical accuracy of axial motion magnification when amplifying the motions, which move in various directions, into only the user-defined direction.
As shown in Fig.~\ref{figure:klt_analysis}-(a), utilizing the Kanade-Lucas-Tomasi (KLT) Tracker~\cite{lucas1981iterative}, we obtain the displacements of the original video and
the video obtained from our method which is magnified
$20$
times along the $x$-axis.
We evaluate both the physical accuracy and efficacy of axial motion magnification by comparing the displacement values from the trajectory of the video amplified $20\times$ using our method against the displacement values obtained by multiplying the original video's displacement values by $20$.
Figure.~\ref{figure:klt_analysis}-(b) demonstrates the alignment between the trajectories of the video obtained by our method and the amplified original trajectory.
For the y-axis displacement, the direction our method does not aim to amplify, the trajectory of the video obtained by our method aligns with the original trajectory. (Fig.~\ref{figure:klt_analysis}-(c)).
These results demonstrate that the proposed axial motion magnification not only preserves physical accuracy but also selectively amplifies motion along user-defined directions.

\begin{figure}[!t]
\centering
\includegraphics[width=0.85\linewidth]{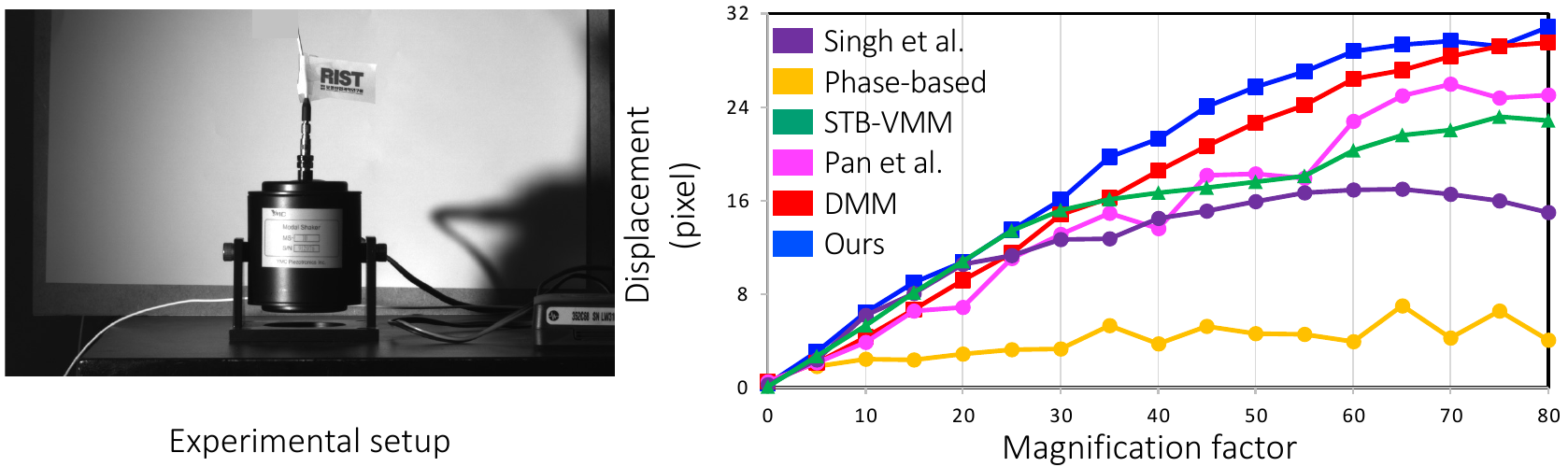}
    \caption{
    \textbf{Amplified motion magnitude with varying magnification factors.} The phase-based method, which is based on signal processing, reached its amplification limit the quickest. Among the learning-based methods, both our method and DMM~\cite{oh2018learning} showed the most proportional results between magnification factor and displacement. However, these methods also reach their amplification limit at large magnification factors, specifically those above 60.
    }\label{figure:mag_limit}
\end{figure}

\subsection{Analysis of Amplifiable Motion Magnitudes}\label{sec:B.5}
The Wu~\etal~\cite{wu2012eulerian} and phase-based method~\cite{wadhwa2013phase}, which are based on signal processing techniques, showed that the amplifiable motion magnitude is theoretically limited. DMM~\cite{oh2018learning}, which is the learning-based method, discussed that the magnification is bounded by the effective receptive field size of the decoder. Our model is developed based on DMM, so it is also subject to this magnification bound.
To demonstrate this experimentally, we amplify the motion of vibrator video using motion magnification methods~\cite{wadhwa2013phase,oh2018learning,lado2023stb,singh2023multi,pan2024self} and measured the amplitude of the vibrator using KLT. The vibrator used in this experiment is the same as the one used in Sec.~\ref{sec:B.3}, but in this experiment, we changed the focal length of the camera lens to $50$mm. As shown in Fig.~\ref{figure:mag_limit}, the phase-based method reaches its amplification limit with smaller motion magnitudes. In the learning-based methods, both DMM and our method show the most proportional results between magnification factor and displacement. However, these methods also reach their amplification limit at large magnification factors, specifically those above 60.

\begin{figure}[t]
\includegraphics[width=0.92\linewidth]{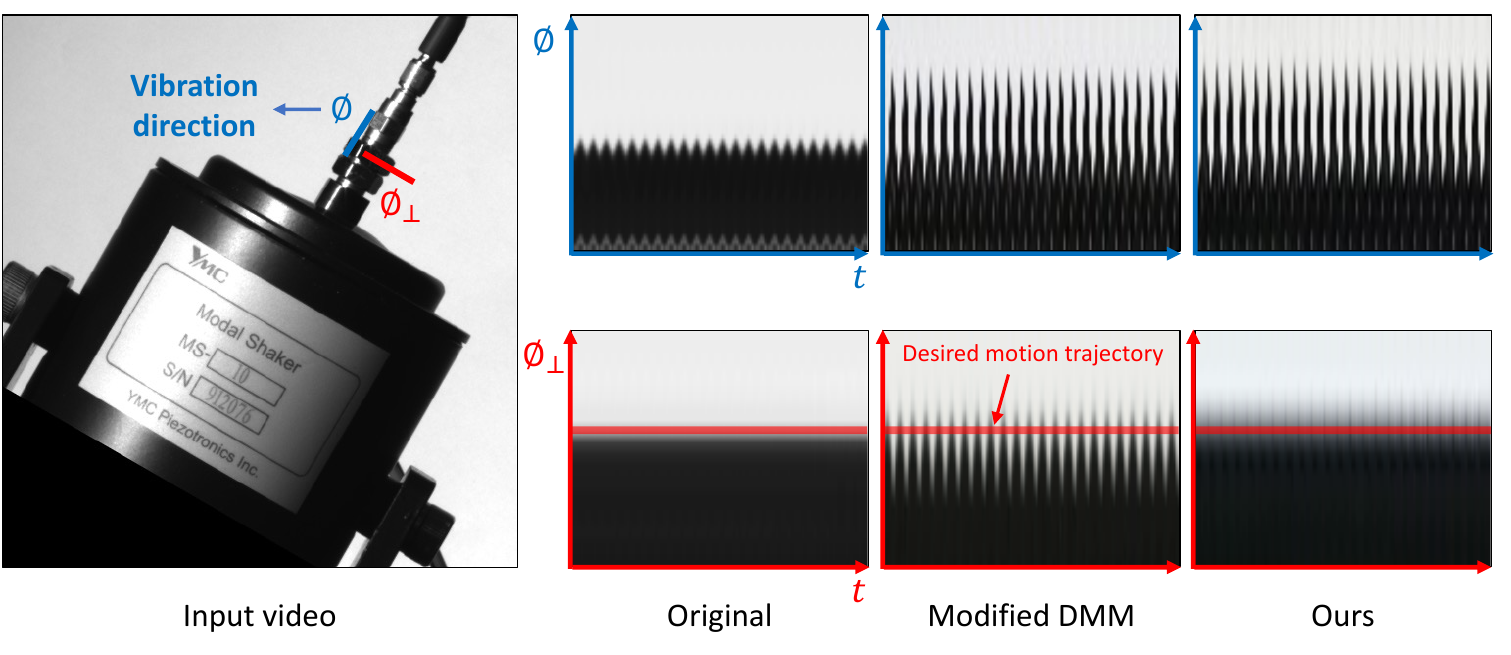}
\centering
    \caption{
    \textbf{Motion separation experiment with MSM.} Along the $\phi_\perp$ direction, we apply the $10\times$ axial motion magnification to the video of a vibrator oscillating only in the $\phi$ direction, using both our method and the modified DMM.
    Contrary to
    the $\phi_\perp$-t slice of the original, the modified DMM exhibits vibration in the $\phi_\perp$ direction due to the unsuccessful motion separation.
    In comparison, our method, leveraging the proposed Motion Separation Module (MSM), successfully distinguishes between the two orthogonal motions, resulting in a $\phi_\perp$-t slice that closely resembles the original's and desired motion trajectory, demonstrating the effectiveness of the MSM.
    }
    \label{figure:MSM}
\end{figure}

\subsection{Motion Separation Effect of the MSM}\label{sec:B.6}
We assess the effectiveness of the Motion Separation Module (MSM) in distinguishing between two orthogonal directional motions.
To explore this, we rotate the video,
where
a vibrator
oscillates
solely along the $y$-axis, by the angle $\phi$ and apply the $10\times$ axial motion magnification to the video along the $\phi_\perp$ direction using both our method and the modified DMM with the $\textit{static}$ mode.
Subsequently, we compare the time slices in the direction of $\phi_\perp$, i.e., the direction with no motion.
In this experiment, we set $\phi$ to $30$ degrees. Figure~\ref{figure:MSM} demonstrates the results.
Unlike the $\phi_\perp$-t slice of the original, where there is no motion in the $\phi_\perp$ direction, modified DMM fails to separate the motion and exhibits motion in the $\phi_\perp$ direction. In contrast, Ours with MSM effectively separates the motions in two orthogonal directions, showing results similar to the original in the $\phi_\perp$ direction.

\begin{figure}[t]
\includegraphics[width=0.92\linewidth]{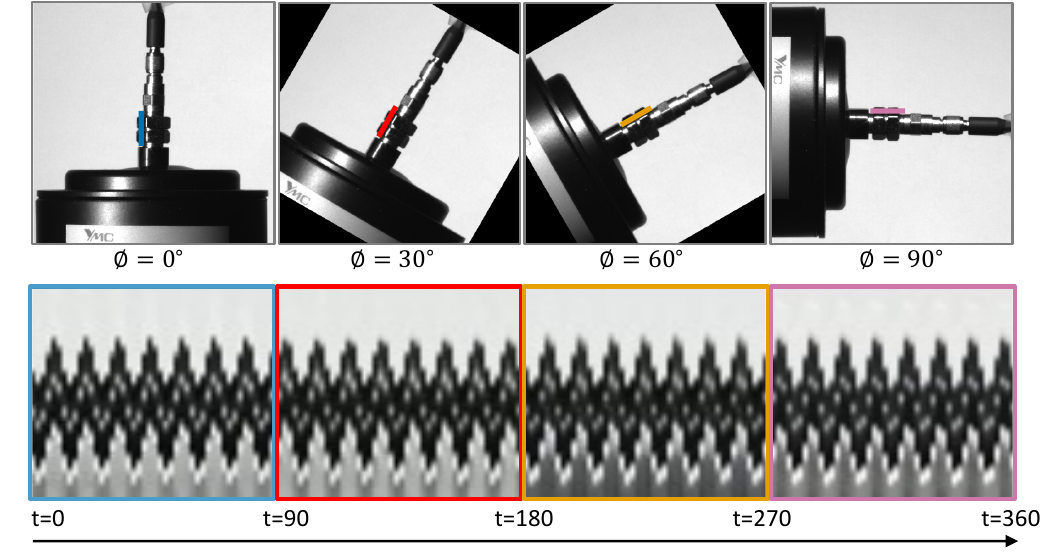}
\centering
    \caption{
    \textbf{Axial motion magnification results across various angles.} We apply axial motion magnification to amplify motion in the direction $\phi$ for vibrator videos rotated at various angles $\phi$. The time slices of axially amplified videos using our method show the smooth transition at boundaries where the angle $\phi$ changes.
    }
    \label{figure:rot_inv_sup}
\end{figure}

\subsection{Angular Analysis of Axial Motion Magnification}\label{sec:B.7}
Our learning-based axial motion magnification can magnify the motion along the user-defined direction.
We examine whether the behavior of our learning-based axial motion magnification remains consistent with changing angles.
As shown in Fig.~\ref{figure:rot_inv_sup}, we rotate the vibrator video at various angles $\phi$ and apply $10\times$ axial motion magnification to amplify only the motion corresponding to $\phi$.
Time slices are obtained from the lines that indicate identical positions across the various angle-adjusted videos.
Then, the slices are sequentially connected over time.
The connected time slices exhibit a smooth transition at boundaries where the angle $\phi$ changes.
This demonstrates the consistent behavior of our learning-based axial motion magnification across various angles.

\begin{figure}[t]
\includegraphics[width=1\linewidth]{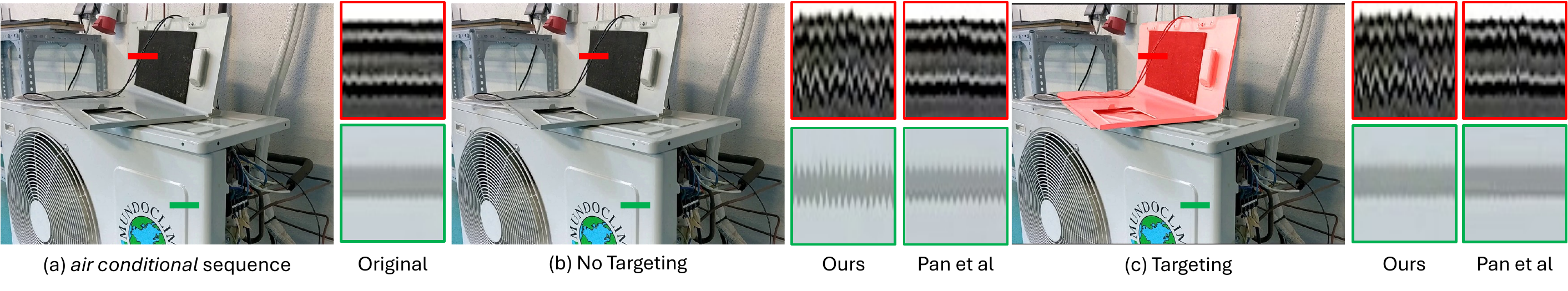}    \caption{
    \textbf{Qualitative results of targeted motion magnification.} Our method is capable of per-pixel motion magnification because of the new proposed training dataset. To show this, Given a mask of the notebook, we selectively magnify the motion of the notebook using Ours and Pan~\etal~\cite{pan2024self}. When targeting the notebook for magnification, we observe that the motion of the air conditioner remains unchanged from the original, while only the motion of the notebook is amplified in Pan~\etal and Ours.
    }\label{figure:targetedmm}
\end{figure}

\subsection{Per-pixel Motion Magnification}\label{sec:B.8}
During inference, our model demonstrates the ability to perform per-pixel motion magnification, which enables us to vary magnification factors across different areas within an image.
This capability is endowed by two main components: the angle $\phi$ and object-wise magnification map $\Lambda$, which are the main parts of our newly proposed training dataset.
We show this
spatially selective
motion magnification capability by presenting targeted results similar to those achieved by Pan~\etal~\cite{pan2024self}, which magnify specific objects within an image.
Figure~\ref{figure:targetedmm} displays the targeted motion magnification results of our method, alongside the targeted results obtained by Pan~\etal
When focusing on magnifying the motion of a notebook, we observe that the motion of the air conditioner remains unchanged from the original footage, while only the motion of the notebook is magnified in both Pan~\etal and our method.

\begin{table}[t]
    \centering
    \caption{\textbf{Comparison of inference speed.}
    Using a 720p resolution video, we measure the Frame Per Second (FPS) of learning-based motion magnification methods. Singh~\etal show differences in inference speed depending on the model. Pan~\etal find that as the magnification factor increased, speed differences occurred due to the recursive magnification process. STB-VMM exhibited slow inference speed, while ours showed slightly slower speed compared to DMM because of lacking parallel processing of the shape branch and manipulator for each axis.
           }
    \resizebox{0.97\linewidth}{!}{
    \begin{tabular}{>{\centering\arraybackslash}p{1.8cm}>{\centering\arraybackslash}p{1.8cm}>{\centering\arraybackslash}p{1.8cm}>{\centering\arraybackslash}p{1.8cm}>{\centering\arraybackslash}p{1.8cm}>{\centering\arraybackslash}p{2.3cm}>{\centering\arraybackslash}p{2.0cm}>{\centering\arraybackslash}p{1.0cm}}
        \toprule
         \multirow{2}{*}{Method} & \multicolumn{2}{c}{Singh~\etal~\cite{singh2023multi}} &  \multicolumn{2}{c}{Pan~\etal~\cite{pan2024self}} & \multirow{2}{*}{STB-VMM~\cite{lado2023stb}} &  \multirow{2}{*}{DMM~\cite{oh2018learning}} & \multirow{2}{*}{Ours} \\
         \cmidrule(lr){2-3} \cmidrule(lr){4-5}
         & D1-model & D2-model & 1-recursive & 2-recursive &  & \\
         \midrule
        FPS & 8.4 & 7.2 & 19.6  & 9.8 & 2.6 & 17.4 & 14.4 \\
        \bottomrule
    \end{tabular}}
    \label{tab:inference_speed}
\end{table}

\subsection{Comparison of Inference Speed}\label{sec:B.9}
We report the inference speed of our method and other learning-based motion magnification methods, including DMM, STB-VMM, Singh~\etal, and Pan~\etal
We measure the inference speed at a 720p resolution video using a single NVIDIA RTX A6000 GPU, which is different from the two NVIDIA TITAN RTX GPUs used during training. Table~\ref{tab:inference_speed} summarizes the results. Singh~\etal proposed the D1 and D2 models, which exhibited inference speeds of 8.4 FPS and 7.2 FPS, respectively. Pan~\etal perform motion magnification recursively if the magnification factor exceeds a certain threshold. In the publicly available implementation,recursive magnification is applied twice when the magnification factor is 16 or higher, which cuts the inference speed in half.
STB-VMM adopts the transformer architecture and shows slow inference speed, as described in the limitations of its own paper.
DMM shows 17.4 FPS, while our method achieves 14.4 FPS. This difference is due to the fact that our implementation is currently serial, lacking parallel processing of the shape branch and manipulator for each axis, which causes computational bottlenecks. Implementing parallel processing for these modules could enhance the inference speed of our method.

\section{Additional Results on Diverse Scenarios}\label{sec:C}
\begin{figure}[t]
\includegraphics[width=1\linewidth]{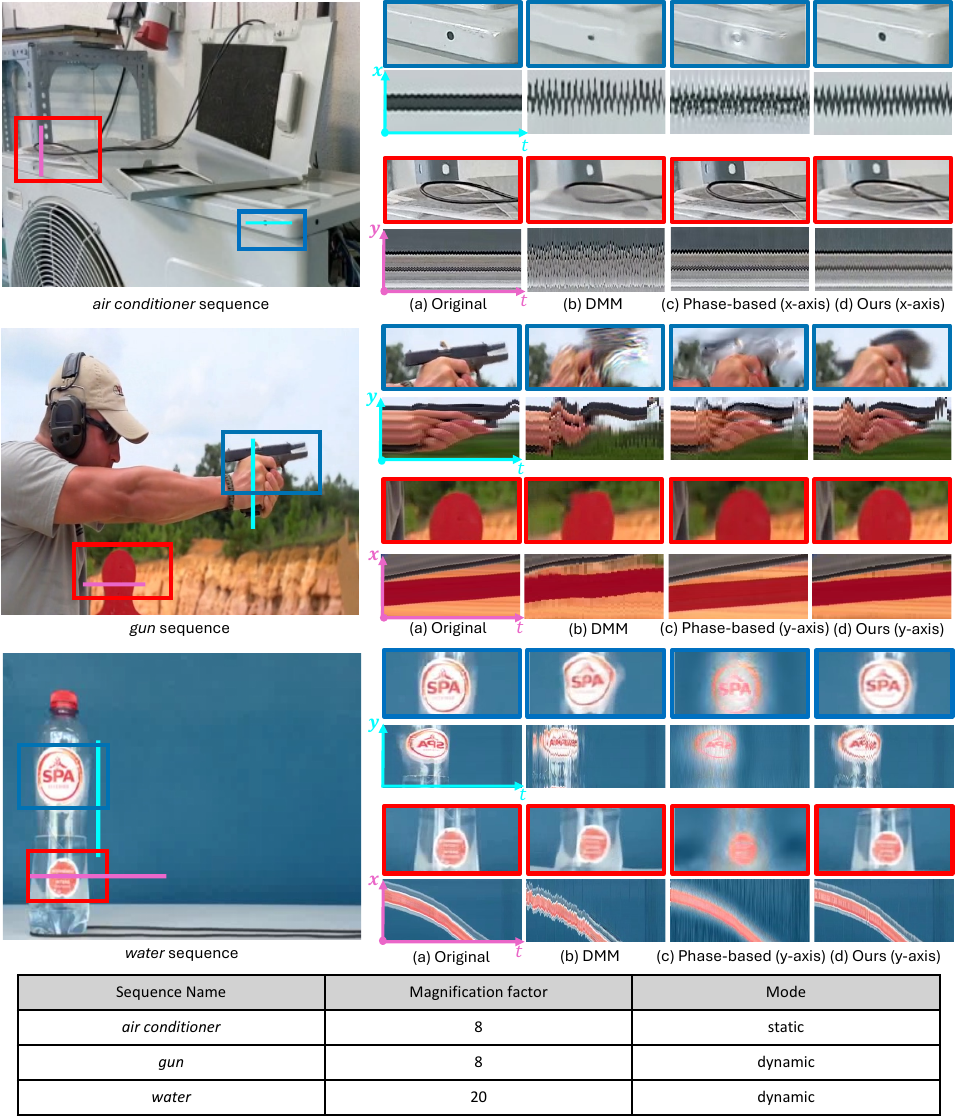}
    \caption{
    \textbf{Qualitative results of axial magnification.
    } (a)~Original: non-magnified. (b)~DMM: magnified results with \textit{generic} method~\cite{oh2018learning}. We magnify $x$-axis motions in \textit{air conditioner} and $y$-axis motions in \textit{gun}, and \textit{water} with (c)~phase-based and (d)~our methods respectively, plotting $x$-$t$ and $y$-$t$ slices for each of two different points.
    In \textcolor{cyan}{cyan} scenarios, where magnification aligns with the slice’s axis, ours presents fewer artifacts and clearer axial vibrations than phase-based, which suffers from severe artifacts and unclear vibrations.
    In \textcolor{magenta}{magenta} scenarios, when magnification is orthogonal to the slice's axis, our method isolates motion effectively, preserving time slices similar to (a) without undesired magnification or artifacts. Conversely, DMM and phase-based struggle, causing time slices to deviate from the original and resulting in notable artifacts.
    }\label{figure:axial_qual}
\end{figure}

\begin{figure}[t]
\includegraphics[width=1\linewidth]{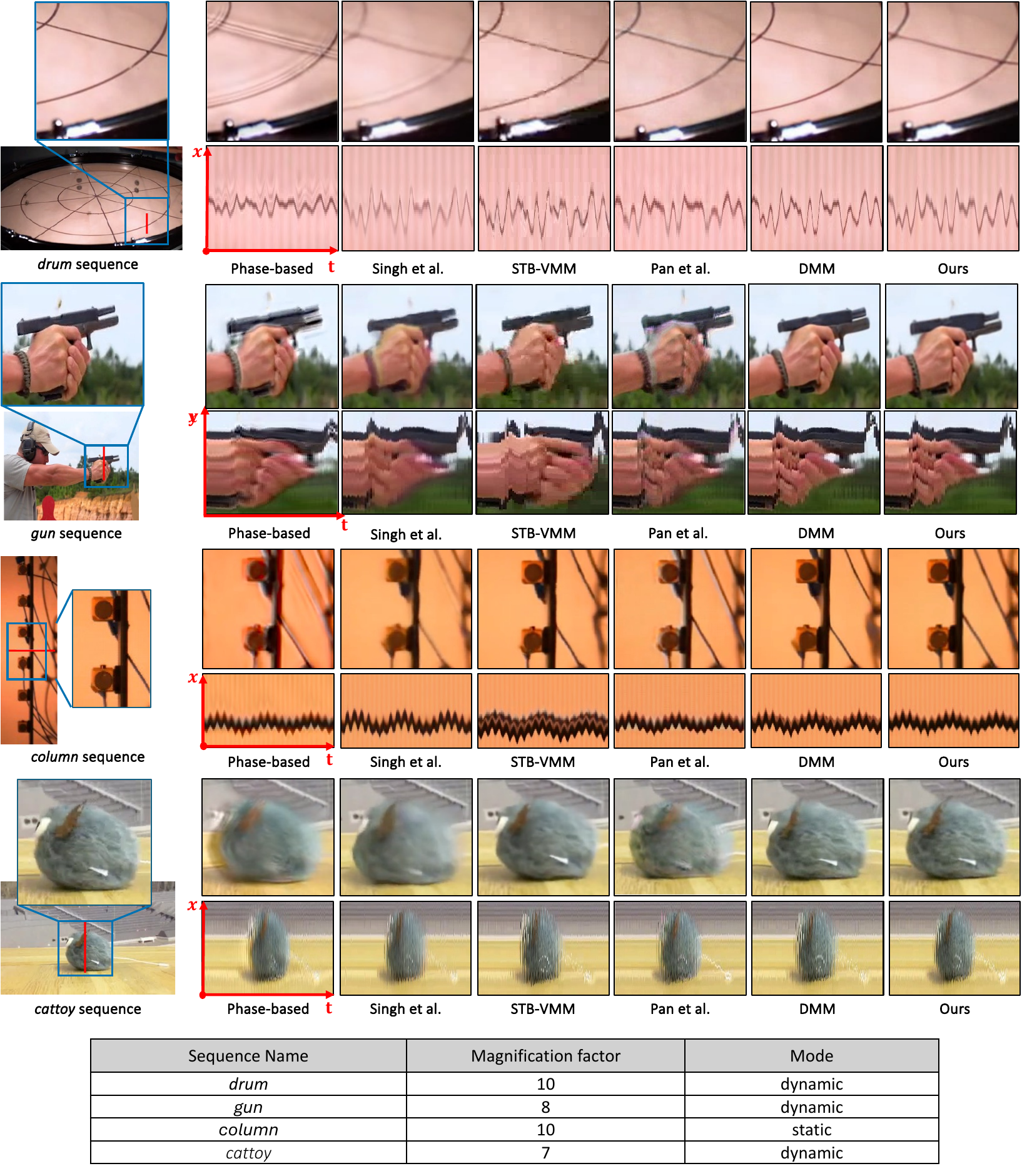}
\centering
    \caption{
    \textbf{Qualitative results of generic motion magnification.} We compare our method to the phase-based~\cite{wadhwa2013phase} method, Singh~\etal~\cite{singh2023multi}, STB-VMM~\cite{lado2023stb}, Pan~\etal~\cite{pan2024self}, and DMM~\cite{oh2018learning} in general motion magnification across various scenarios. Our method demonstrates clear magnified frames and the $x$-t slices.
    }
    \label{figure:generic_nofilter}
\end{figure}

\begin{figure}[t]
\includegraphics[width=1\linewidth]{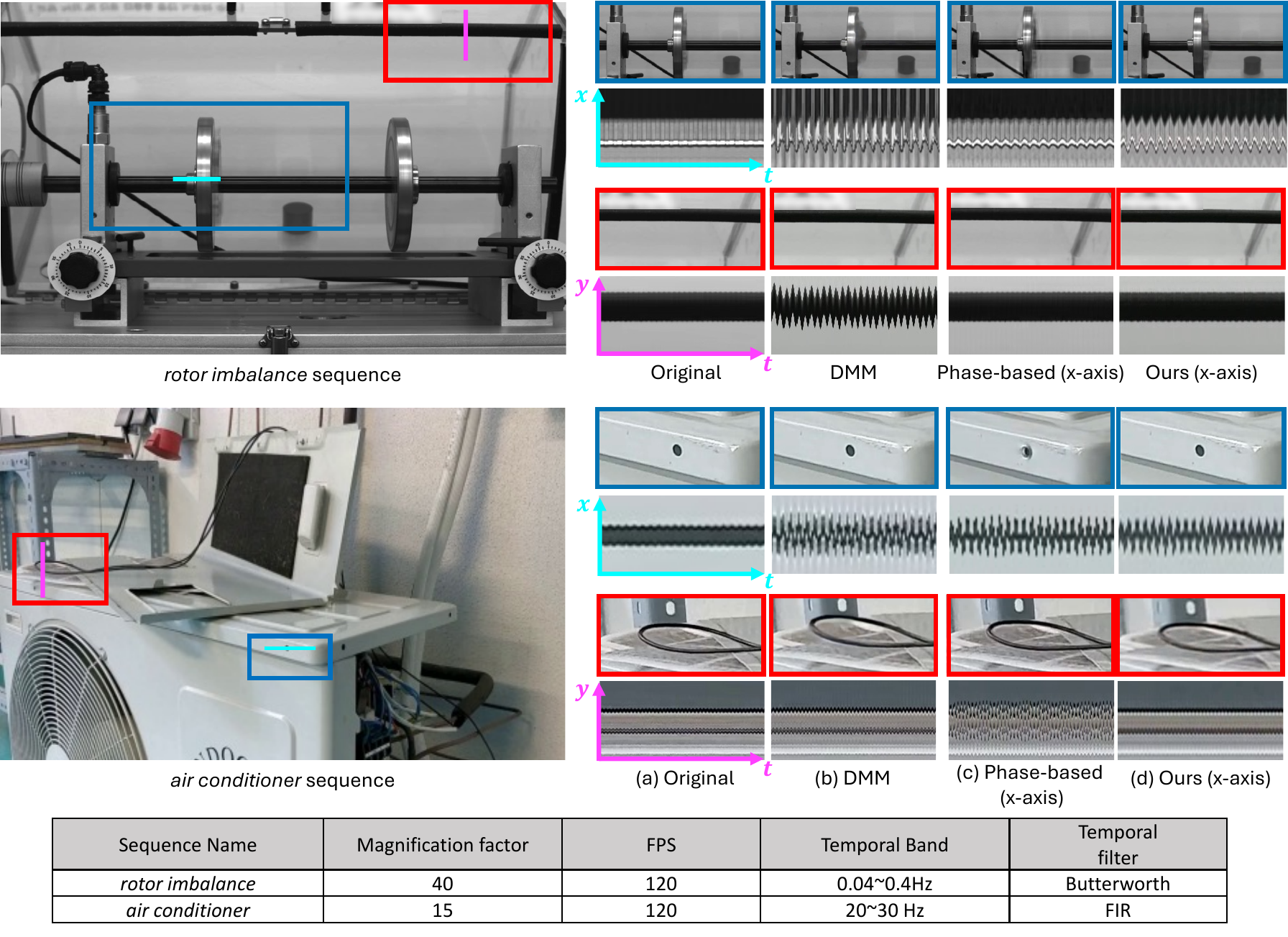}
\centering
    \caption{
    \textbf{Axial motion magnification with temporal filter.}  With the temporal filters, we magnify the \textit{rotor imbalance} and \textit{air conditioner} sequences along the $x$-axis, \ie, the axial direction, using (d) Ours and (c) phase-based method~\cite{wadhwa2013phase}.
    We also show the result of (b) DMM~\cite{oh2018learning} with the temporal filter as one reference result of \textit{generic} motion magnification methods.
    In \textcolor{cyan}{cyan} scenarios, where magnification aligns with the slice’s axis, ours shows fewer artifacts and legible axial vibrations.
    On the other hand, DMM and phase-based methods suffer from severe artifacts.
    In addition, DMM shows unclear vibration in the $x$-t slice, even with the temporal filter.
    In \textcolor{magenta}{magenta} scenarios, when magnification is orthogonal to the slice's axis, our method effectively isolates the motions that are not aligned with the magnified direction, preserving time slices similar to (a) Original without undesired magnification or artifacts.
    Conversely, DMM in rotor imbalance sequence
    and phase-based in air conditioner sequence
    struggle to disentangle the unwanted motions, which leads to time slices deviating from the original and the magnified frames with artifacts and unclear axial vibrations.
    }
    \label{figure:temporal_axial_aa}
\end{figure}

\begin{figure}[t]
\includegraphics[width=1\linewidth]{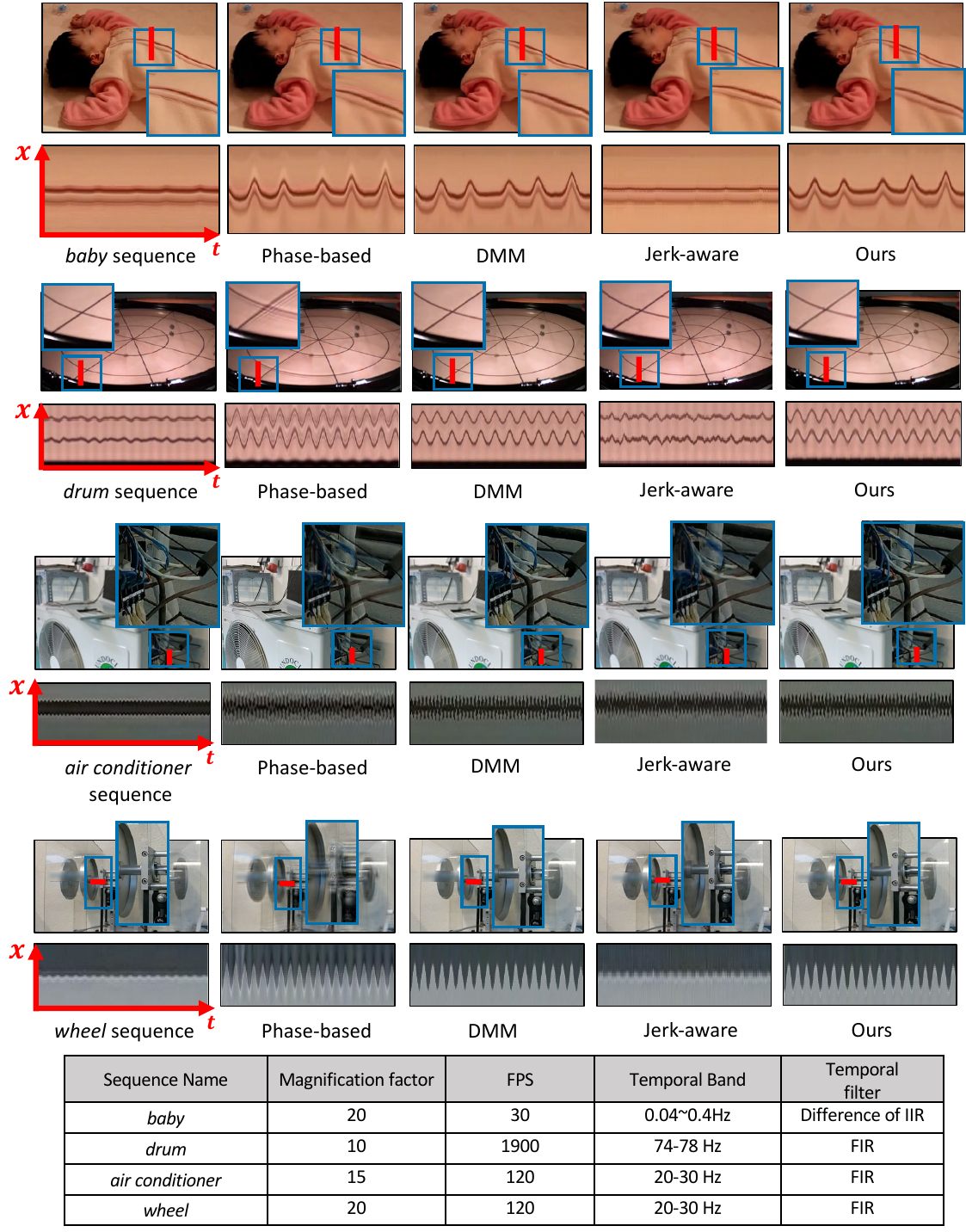}
\centering
    \caption{
    \textbf{\textit{Generic} motion magnification with temporal filter.} With temporal filters, we applied \textit{generic} motion magnification to the \textit{baby}, \textit{drum}, \textit{air conditioner}, and \textit{wheel} sequence using the phase-based method, DMM~\cite{oh2018learning}, Jerk-aware~\cite{takeda2018jerk} and our methods. Ours and DMM preserve the boundaries of the moving objects while depicting the motion well. The phase-based method exhibits slight ringing artifacts, and the Jerk-aware method shows the unstable separation of the motion signals.
    }
    \label{figure:temporal_generic}
\end{figure}

In this section, to demonstrate the efficacy of our approach, we present results from diverse scenarios.
Our method is capable of both generic and axial motion magnification. Additionally, we observe that the learned shape representations are compatible with the temporal filter, similar to DMM~\cite{oh2018learning}.
Therefore, our proposed method provides four configurations based on the motion magnification approach and the application of temporal filters.
The following figures demonstrate results on four distinct configurations: axial motion magnification without a temporal filter (Fig.~\ref{figure:axial_qual}), generic motion magnification without a temporal filter (Fig.~\ref{figure:generic_nofilter}), axial motion magnification with a temporal filter (Fig.~\ref{figure:temporal_axial_aa}), and generic motion magnification with a temporal filter (Fig.~\ref{figure:temporal_generic}).

\end{document}